%% file: main.tex
 \documentclass[10pt,onecolumn,twoside]{IEEEtran} 
 
\usepackage{amsmath,amsopn,amssymb}
\usepackage{graphicx,xspace,color}
\usepackage{epsfig,subfigure}
\usepackage{longt able,multirow}
\usepackage{array,float}
\usepackage{cite}
\usepackage[nolist]{acronym}
\usepackage{soul}

\newif\ifimportant
\importantfalse 

\begin{document}

\begin{acronym}
\acro{DIBR}{depth-image based rendering}  
\acro{DP}{dynamic programming}  
\acro{RTT}{round-trip time}
\acro{SC}{set covering}  
\acro{IMVS}{interactive multiview video system}  
\acro{MRF}{Markov random field}
\end{acronym}

\title{In-Network View Synthesis for Interactive Multiview Video Systems}

\author{
Laura Toni~\IEEEmembership{Member,~IEEE},
Gene Cheung~\IEEEmembership{Senior Member,~IEEE}, and
Pascal Frossard~\IEEEmembership{Senior Member,~IEEE}
\begin{small}
\thanks{L. Toni, and P. Frossard are with \'Ecole Polytechnique F\'ed\'erale de Lausanne (EPFL), Signal Processing Laboratory - LTS4, CH-1015 Lausanne, Switzerland. Email: \texttt{\{laura.toni,  pascal.frossard\}@epfl.ch.} }
\thanks{Gene Cheung is with the National Institute of Informatics, Tokyo, Japan. Email Address: \texttt{cheung@nii.ac.jp}}
\end{small}
}%
\maketitle
 
\begin{abstract}
Interactive multiview video applications endow users with the freedom to navigate  through neighboring  viewpoints in a 3D scene. To enable such interactive navigation with a minimum view-switching delay, multiple camera views are sent to the users, which are used as reference images to   synthesize additional    virtual  views via depth-image-based rendering. In practice, bandwidth constraints may however restrict the number of reference views sent to clients per time unit, which may in turn limit the quality of the synthesized viewpoints. 
We argue that the reference view selection should ideally be performed close to the users, and we study the problem of in-network reference view synthesis such that the navigation quality is maximized at the clients.  We consider a distributed  cloud  network architecture   where  data  stored in a main cloud  is delivered to end users with the help of  cloudlets, i.e., resource-rich proxies close to the  users.   
In order to satisfy last-hop bandwidth constraints from the    cloudlet   to the users,  a  cloudlet \emph{re-samples} viewpoints of the 3D scene into a discrete set of views (combination of received camera views and virtual views synthesized)  
  to be used as reference for the synthesis of additional virtual views at the client. This in-network synthesis leads to better viewpoint sampling given a bandwidth constraint compared to simple selection of camera views, but it may however carry a distortion penalty in the cloudlet-synthesized reference views.    We therefore cast a new reference view selection problem where the best subset of views is defined as the one minimizing the distortion over a view navigation window defined by the user  under some transmission bandwidth constraints. We show that the view selection problem is NP-hard, and propose an effective  polynomial time algorithm using dynamic programming to solve the optimization problem under general assumptions that cover most of the multiview scenarios in practice.  Simulation results finally  confirm the  performance gain offered by  virtual view synthesis in the network. It shows that cloud computing resources provide important benefits in resource greedy applications such as interactive multiview video.
\end{abstract}


\IEEEpeerreviewmaketitle

\section{Introduction}
\label{sec:intro}
\input{intro_new}

\section{Related Work}
\label{sec:related}
\input{related}

\section{Background}
\label{sec:sys_overview}
\input{system_overview}

\section{Reference View Selection Problem}
\label{sec:algo}
\input{model}
\section{Optimal View Selection Algorithm}
\label{sec:algo_approx}
\input{algo_approx}
\section{Simulation  Results}
\label{sec:results}
\input{results}


\section{Conclusion}
\label{sec:conclusion}
When interactive multiview video systems face  limited bandwidth constraints, we argue that synthesizing   reference views in the cloud improve the quality of navigation at the client side.  In particular, we propose  a synthesized reference view selection optimization problem aimed at finding the best subset of  viewpoints to be transmitted to the decoder as reference views. This subset is not limited to captured camera views as in previous approaches but it can also include virtual viewpoints. The problem  is formalized  as a combinatorial optimization problem, which is shown to be  NP-hard.  However, we show that,  under the general assumption that the distortion of synthesized viewpoints  is well-behaved, the problem can be solved in polynomial time via a dynamic programming algorithm. Simulation results validate the performance gain of the proposed method and   show that synthesizing reference views can improve image quality at the client by up to $2.1$dB in PSNR.  We    finally demonstrate that view synthesis  in the network   obviates to non optimal camera sampling and permits to increase the distance between camera views without affecting the quality  of the navigation.




\bibliographystyle{IEEEbib}
\bibliography{ref,MVref}

\end{document}

%% file: intro_new.tex
Interactive free viewpoint video systems \cite{tanimoto11} endow users with the ability to choose and display any virtual view of a 3D scene, given original viewpoint images captured by  multiple cameras. In particular, a virtual view image can be synthesized by the decoder via \textit{depth-image-based rendering} (DIBR)~\cite{Fehn_C_2004} using texture and depth images of two neighboring views that act as reference viewpoints. 
One of the key challenges in \textit{interactive multiview video streaming} (IMVS)  \cite{Gene:J11} systems    is to transmit an appropriate subset of reference views from a potentially large number of camera-captured views such that the client enjoys   high quality and low   delay view navigation even in resource-constrained environments~\cite{LiuQinMaZha:J10,  Chack:J13,Toni:J14}.  
%
  %

\begin{figure}[t]
\begin{center}
\includegraphics[width=0.9\linewidth,draft=false]{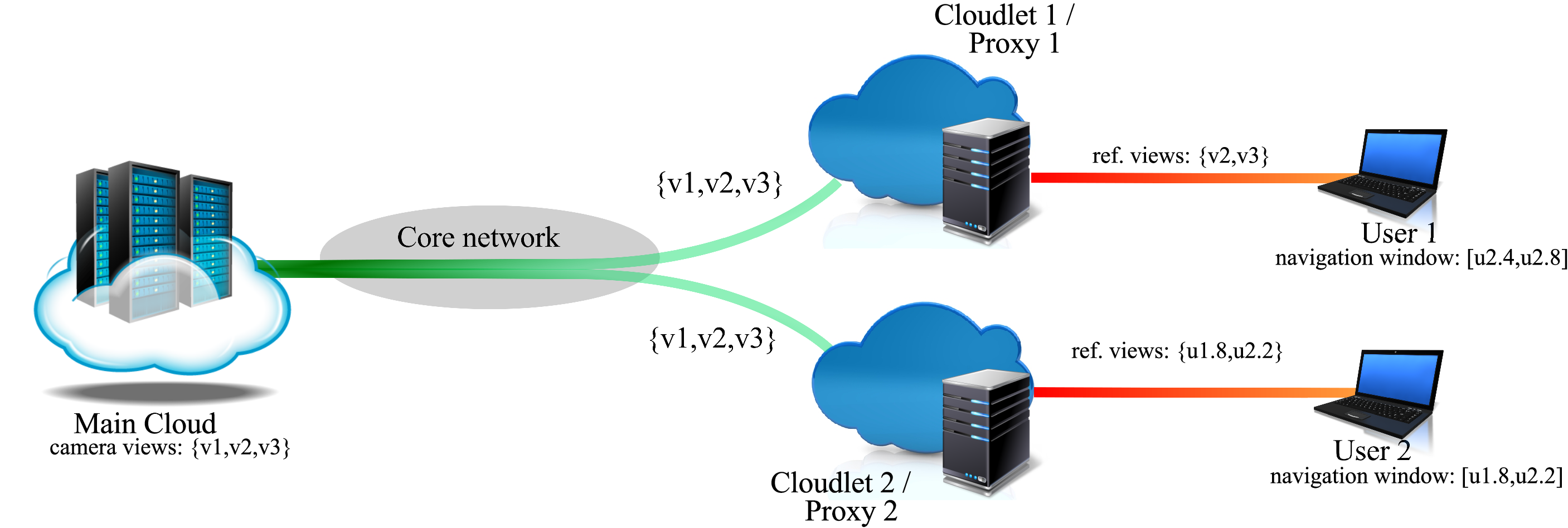}
\caption{Considered scenario. Green lines represent no bandwidth constrained channels, red lines are bottleneck channels.}
\label{fig:ConsideredScenario}
\end{center}
\end{figure}

In this paper, we propose a new paradigm to solve the reference view selection problem and capitalize on cloud computing resources to perform fine adaptation close to the clients. 
We consider a hierarchical cloud framework,    where the   selection of reference views is performed by  a \emph{network of cloudlets}, i.e.,  resource-rich proxies that can perform personalized processing at the edges of the core network \cite{Verbelen:2012,Satyanarayanan:J09}. 
An adaptation at the cloudlets  results in  a smaller \ac{RTT}, hence more reactivity than in more centralized architectures.   
Specifically, we consider the scenario depicted  in Fig.~\ref{fig:ConsideredScenario}, where a main cloud   stores  pre-encoded    video from different cameras, which are then  transmitted to the 
edge cloudlets  that act as proxies for final delivery to users. 
We assume that  there is sufficient network capacity between the main cloud and the edge cloudlets for the transmission of all   camera views, but there exists however a bottleneck of limited capacity  between a cloudlet and a nearby user\footnote{In practice, the last-mile access network is often the bottleneck in real-time media distribution.}.  In this  scenario, each cloudlet sends to a client the set of reference views that respect bandwidth capacities and enable synthesis of all viewpoints in the client's navigation window. This window is defined  as the range of viewpoints in which the user can navigate during the RTT and enables   zero-delay   view-switching at the client.

We argue that, in  resource-constrained networks, \textit{re-sampling the viewpoints of the 3D scene} in the network--- i.e., synthesizing novel virtual views in the cloudlets that are transmitted as new references to the decoder---is beneficial compared to the mere subsampling of the original set of camera views.   We illustrate this in   Fig.~\ref{fig:ConsideredScenario}, where the main cloud  stores three coded camera views: $\{v_1, v_2, v_3\}$ while  the bottleneck links between cloudlet-user pairs can support the transmission of only two views.    \footnote{We consider integer index $i$ for any    camera view,  while  we assume that a virtual view can have a non-integer index $i.x$, which corresponds to a position between camera views $v_i$ and $v_{i+1}$. }   If user 1 requests a navigation window $[u_{2.4}, u_{2.8}]$, the   cloudlet can simply forward the closest camera views $v_2$ and $v_3$. However, if user 2 requests the navigation window $[u_{1.8}, u_{2.2}]$, transmitting camera views $v_1$ and $v_3$ results in large synthesized view distortions due to the large distance between reference and virtual views (called \textit{reference view distance} in the sequel). Instead, the cloudlet can synthesize virtual views $u_{1.8}$ and $u_{2.2}$ using camera views $v_1, v_2, v_3$ and send these virtual views to the user 2 as new reference views for the navigation window $[u_{1.8}, u_{2.2}]$. This strategy may result in smaller synthesized view distortion due to the smaller distance to the reference views. However, the in-network virtual view synthesis may also introduce distortion into the new reference views $u_{1.8}$ and $u_{2.2}$, which results in a tradeoff that should be    carefully considered when choosing the views to be synthesized in the cloudlet. 
 
Equipped with the above intuitions, we study the main tradeoff between reference distortion and bandwidth gain. Using a Gauss-Markov model, we first analyze the benefit of synthesizing new reference images in the network. We then formulate a new   \emph{synthesized reference view selection optimization problem}.  It consists in  selecting  or constructing the optimal  reference views  that lead to the minimum distortion for all synthesized virtual views in the user's navigation window subject  to a bandwidth constraint between the cloudlet and the user.   
We show that this combinatorial problem can be  solved optimally but that it is NP-hard. We then introduce a generic assumption on the view synthesis distortion  which leads to a  polynomial time solution with 
a \ac{DP} algorithm. 
   We then provide  extensive simulation results  for synthetic and natural sequences. They confirm   the quality gain experienced by the IMVS clients  when synthesis is allowed in the network,  with respect to scenarios whose edge cloudlets can only transmit camera views.  They also show that synthesis in the network allows to maintain good navigation quality when reducing the number of   cameras as well as when  cameras are not ideally positioned in the 3D scene. 
This  is an important advantage in practical settings, which confirms that cloud processing resources can be judiciously used to improve the performance of applications that are a priori quite greedy in terms of network resources.   

The remainder of this paper is organized as follows. Related works are described in Section \ref{sec:related}. In Section \ref{sec:sys_overview}, we provide a system overview and analyze the benefit of in-network view synthesis via a Gauss-Markov model to impart intuitions. The reference view selection optimization problem is then formulated in Section \ref{sec:algo}.
We propose general assumptions on view synthesis distortion in Section \ref{sec:algo_approx} and derive an additional    polynomial time view selection  algorithm. In Section \ref{sec:results}, we discuss the simulation results, and we conclude in Section \ref{sec:conclusion}.

%% file: related.tex
 
Prior  studies    addressed  the problem of providing interactivity in selecting views in IMVS, while saving on transmitted bandwidth and view-switching delay~\cite{Merkle:J07,salvador:J13,Maug:Arxiv14,Fuji:J14,CheOrtChe:C09,Gene:J11, xiu12,DeAbreu:L13}. These works are mainly focused on optimizing the frame coding structure to improve interactive media services. 
In the case of pre-stored camera views, however, rather than optimal frame coding structures,   
 interactivity in network-constrained scenario can be addressed  by studying  optimal camera selection strategies, where a subset of selected camera views is actually transmitted to clients such that the navigation quality is maximized and resource constraints are satisfied \cite{LiuQinMaZha:J10,PanIkuBanWat:C11,Chack:J13,ren15,Toni:J14,Maugey:C14}.  In  \cite{Toni:A14}, an optimal camera view selection algorithm  in resource-constrained networks has been proposed based   on the users' navigation paths.   In \cite{cheung11tip2}   a  bit allocation algorithm over an optimal subset of camera views is proposed for optimizing  the visual distortion of reconstructed views in interactive systems.   
Finally,  in \cite{Toni:C13,Abreu:A15}  authors optimally organize camera views  into layered subsets that are  coded and delivered to clients  in  a prioritized fashion to accommodates for the network and clients heterogeneity and to effectively exploit the resources of the overlay network. While in these works  the selection is limited to camera views, in our work we rather assume in-network processing able to synthesize  virtual viewpoints in the cloud network.  

In-network adaptation strategies  allow to   cope with network resource constraints and are   mainly categorized in $i)$ packet-level processing  and  $ii)$  modification of the source information. In the first  category,  packet filtering,  routing strategies   \cite{De_Vleeschouwer:J07, Huang:C12}  or  caching of media content information \cite{Borst:C10,JiCaire:arxiv13}  allow to save network resources while improving the quality experienced by clients. To better address media delivery services in highly heterogenous scenarios,   network coding strategies  for multimedia streaming have been also proposed~\cite{Dimakis:J10,Magli:J13,Vukob:J14}.  In the second category --- in-network processing at the source level ---    the main objective is usually to avoid transmitting large amounts of raw streams to the clients  by processing the source data in the network to reduce both the communication volume and the processing required at the client side.  Transcoding strategies might be collaboratively  performed in peer-to-peer networks \cite{Wu_Huang:J11}  or in the cloud \cite{wen:J14}. Furthermore, source data can be compressed in the cloud \cite{Wu_Huang:J11,Huang_Mei:C11,Yue:J13} to efficiently address users' requests. 
Rather than media processing  in the main  cloud, 
   offloading  resources to a   cloudlet, i.e., a resource-rich machine in the proximity of the users,    might reduce the transmission latency \cite{Verbelen:2012,Satyanarayanan:J09}. This is beneficial for delay-sensitive / interactive applications \cite{Youngbin13,Tel_Schaar:C14,Jia15}.  
Because of the proximity of cloudlets to users,   cloudlet computing has been under intense investigation for cloud-gaming applications, as shown in \cite{Cai15} and references there in. The above works are mainly focused on multimedia processing, rather than on specific multiview scenarios. However, the use of cloudlets in delay sensitive applications motivates the idea of cloudlet-based view synthesis  for IMVS.

Cloud processing for multiview system is considered in \cite{ Melodia:J13, Xu:J15,Wen:J13}. In \cite{Xu:J15} authors mainly address the cloud-based processing from a security perspective. In \cite{Wen:J13},   view synthesis in the network has been introduced  for cloud networks to offload  clients' terminals (in terms of complexity). The desired view is synthesized in the cloud and then sent directly to clients.   However, only the  view requested by the client is synthesized. This means that either the desired view is a priori known  at the source or a switching delay is experienced by the clients. To the best of our knowledge, none of the work investigating cloud processing have considered the problem of multi-view interactive streaming under network resource constraints.  In our work, we  propose view synthesis in the network mainly to  both overcome uncertainty of users' requests in interactive systems and to cope with limited network resources.


%% file: system_overview.tex
\subsection{System Model}
Let $\mathcal{V} = \{v_1, \ldots, v_N\}$ be the set of the $N$ camera viewpoints captured by the multiview system. For all camera-captured views,  compressed texture and depth maps are stored at the main cloud, with each texture/depth map pair encoded at the same rate using standard video coding tools like H.264\cite{wiegand03} or HEVC\cite{sullivan12}. The possible viewpoints offered to the users are denoted by $\mathcal{U} = \{u_1, u_{1+\delta}, \ldots, u_N\}$. The set $\mathcal{U}$ contains both synthesized views and camera views for navigation between the leftmost and rightmost camera views, $v_1$ and $v_N$. It is equivalent to offering views $u=k\delta$, where $k$ is a positive integer and $\delta$ is a pre-determined fraction that describes the minimum view spacing between neighboring virtual views. 
We consider that any virtual viewpoint $u\in\mathcal{U}$ can be synthesized using a pair of left and right reference view images $v_L$ and $v_R$, $v_L < u < v_R$, via a known DIBR technique such as 3D warping\footnote{Note that view synthesis can be performed in-network (to generate new reference views) or at the user side (to render desired views for observation). In both cases, the same rendering method and distortion model apply. }~\cite{Mori200965}. 
  
Each user is served by an assigned cloudlet through a bottleneck link of capacity $C$, expressed in number of views. Assuming a \ac{RTT} of $T$ seconds between the cloudlet and the user, and a maximum speed $\rho$ at which a user can navigate to neighboring virtual views, one can compute a \textit{navigation window} $W(u) = [u-\rho T,u+\rho T]$, given that the user has selected virtual view $u$ at some time $t_0$. 
The goal of the cloudlet is to serve the user with the best subset of $C$  viewpoints in  $\mathcal{U}$ that synthesize the best quality virtual views in $W(u)$. In this way, the user can experience zero-delay view navigation at time $t_0 + T$ (see \cite{xiu12} for details) with optimized visual quality.
 

 

\subsection{Analysis of Cloudlet-based Synthesized Reference View}

To impart intuition of why synthesizing new references at in-network cloudlets may improve rendered view quality at an end user, we consider a simple model  among neighboring views. Similarly to \cite{li08,zhang13},  we assume a Gauss-Markov model, where variable $x_v$ at view $v$ is correlated with  $x_{v-1}$:
\begin{align}
x_v & = x_{v-1} + e_v, ~~~ \forall v \geq 2
\label{eq:GM}
\end{align}
where $e_v$ is a zero-mean independent Gaussian variable with variance $\sigma_v^2$, and $x_1   = e_1$. A large $\sigma_v^2$ would mean views $x_v$ and $x_{v-1}$ are not similar. We can write $N$ variables $x_1, \ldots, x_N$ in matrix form:
\begin{align}
\mathbf{F} \mathbf{x} = \mathbf{e}, ~~~ \mathbf{x} = \mathbf{F}^{-1} \mathbf{e}
\end{align}
where
\begin{align}
\mathbf{F} = \left[  \begin{array}{ccccc}
1 & 0 & \ldots & & \\
-1 & 1 & 0 & \ldots & \\
0 & -1 & 1 & 0 & \ldots \\
\vdots & & \ddots & \ddots & \\
0 & \ldots & 0 & -1 & 1
\end{array} \right],
~~ \mathbf{x} = \left[  \begin{array}{c}
x_1 \\
\vdots \\
x_N
\end{array}
\right],
~~ \mathbf{e} = \left[  \begin{array}{c}
e_1 \\
\vdots \\
e_N
\end{array}
\right]
\end{align}
Given $\mathbf{x}$ is zero-mean, the \textit{covariance matrix} $\mathbf{C}$ can be computed as:
\begin{align}
\mathbf{C} & = E[\mathbf{x}\mathbf{x}^T] 
 = \mathbf{F}^{-1} E[\mathbf{e} \mathbf{e}^T] (\mathbf{F}^{-1})^T
\end{align}
where $E[\mathbf{e} \mathbf{e}^T] = diag(\sigma_1^2, \ldots, \sigma_N^2)$ is a diagonal matrix. The \textit{precision matrix} $\mathbf{Q}$ is the inverse of $\mathbf{C}$ and can be derived as follows:
\begin{align}
\mathbf{Q} = \mathbf{C}^{-1} 
& = \left( \mathbf{F}^{-1} \; diag(\sigma_1^2, \ldots, \sigma_N^2) \; (\mathbf{F}^{-1} )^T \right)^{-1} \nonumber \\
& = \mathbf{F}^T \; diag(\sigma_1^2, \ldots, \sigma_N^2)^{-1} \; \mathbf{F}
\nonumber \\
& = \left[ \begin{array}{ccccc}
\frac{1}{\sigma_1^2} + \frac{1}{\sigma_2^2} & - \frac{1}{\sigma_2^2} & 0 & \ldots & \\
- \frac{1}{\sigma_2^2} & \frac{1}{\sigma_2^2} + \frac{1}{\sigma_3^2} & - \frac{1}{\sigma_3^2} & 0 & \ldots \\
0 & - \frac{1}{\sigma_3^2} & \frac{1}{\sigma_3^2} + \frac{1}{\sigma_4^2} & - \frac{1}{\sigma_4^2} & \\
\vdots & & \ddots & \ddots & \ddots \\
0 & & & - \frac{1}{\sigma_N^2} & \frac{1}{\sigma_N^2}
\end{array} \right]
\end{align}
which is a tridiagonal matrix. 

When synthesizing a view $x_n$ using its neighbors $x_{n-1}$ and $x_{n+1}$, we would like to know the resulting precision. Without loss of generality, we write $\mathbf{x}$ as a concatenation of two sets of variables, \textit{i.e.} $\mathbf{x} = [\mathbf{y} ~~ \mathbf{z}]$. It can be shown \cite{rule05} that the conditional mean and precision matrix of $\mathbf{y}$ given $\mathbf{z}$ are:
\begin{align}
\mu_{\mathbf{y} | \mathbf{z}} & = 
\mu_{\mathbf{y}} - \mathbf{Q}^{-1}_{\mathbf{y} \mathbf{y}} \mathbf{Q}_{\mathbf{y} \mathbf{z}}
\left( \mathbf{z} - \mu_{\mathbf{z}} \right) \nonumber \\
\mathbf{Q}_{\mathbf{y} | \mathbf{z}} & = \mathbf{Q}_{\mathbf{y} \mathbf{y}}
\label{eq:condPrec}
\end{align}

Consider now a set of four views $x_1, x_2, x_3, x_4$, where $x_1, x_2, x_4$ are camera views transmitted from the main cloud. Suppose further that the user window is $[1.8, 2.2]$, and the cloudlet has to choose between using received $x_4$ as right reference, or synthesizing new reference $x_3$ using received $x_2$ and $x_4$. Using the discussed Gauss-Markov model (\ref{eq:GM}) and the conditionals (\ref{eq:condPrec}), we see that synthesizing $x_3$ using reference $x_2$ and $x_4$ results in precision:
\begin{equation}
Q_{3|(2,4)} = Q_{33} 
= \frac{1}{\sigma_3^2} + \frac{1}{\sigma_4^2}
\end{equation}
$1/Q_{33}$ is thus the additional noise variance when using new reference $x_{\bar{3}}$ to synthesize $x_2$. We can then compute the conditional precision $Q_{2|(1,\bar{3})}$ given new reference $x_{\bar{3}}$:
\begin{equation}
Q_{2|(1,\bar{3})} = \frac{1}{\sigma_2^2} 
+ \frac{1}{\sigma_3^2 + \left( \frac{1}{\sigma_3^2} + \frac{1}{\sigma_4^2} \right)^{-1} }
\label{eq:prec1}
\end{equation}

In comparison, if a user uses received $x_4$ as right reference, $x_4$ will accumulate two noise terms from $x_2$ to $x_4$:
\begin{align}
x_4 = x_2 + e_3 + e_4
\end{align}
The resulting conditional precision of $x_2$ given $x_1$ and $x_4$ is:
\begin{equation}
Q_{2|(1,4)} = \frac{1}{\sigma_2^2} + \frac{1}{\sigma_3^2 + \sigma_4^2}
\label{eq:prec2}
\end{equation}
We now compare $Q_{2|(1,\bar{3})}$ in (\ref{eq:prec1}) with $Q_{2|(1,4)}$ in (\ref{eq:prec2}). We see that if $\sigma_3^2$ is very large relative to $\sigma_4^2$, then $\left( \frac{1}{\sigma_3^2} + \frac{1}{\sigma_4^2} \right)^{-1} \approx \sigma_4^2$, and $Q_{2|(1,\bar{3})} \approx Q_{2|(1,4)}$. That means that if view $x_3$ is very different from $x_2$, then synthesizing new reference $x_3$ does not help improving precision of $x_2$. However, if $\frac{1}{\sigma_3^2} < \infty$, then $\left( \frac{1}{\sigma_3^2} + \frac{1}{\sigma_4^2} \right)^{-1} < \sigma_4^2$, and $Q_{2|(1,\bar{3})} > Q_{2|(1,4)}$, which means that in general it is worth  to synthesize new reference $x_3$. The reason can be interpreted from the derivation above: by synthesizing $x_3$ using both $x_2$ and $x_4$, the uncertainty (variance) for the right reference has been reduced from $\sigma_4^2$ to $\left( \frac{1}{\sigma_3^2} + \frac{1}{\sigma_4^2} \right)^{-1}$,   improving the precision of the subsequent view synthesis.

%% file: model.tex
In this section, we first formalize the synthesized reference view selection problem. We then describe an assumption on the distortion of synthesized viewpoints. We conclude by showing that under the considered assumption the optimization problem is NP-hard.

\subsection{Problem Formulation}

Interactive view navigation means that a user 
 can construct any virtual view within a specified navigation window with zero view-switching delay, using viewpoint images transmitted from the main cloud as reference \cite{xiu12}. We denote this navigation window by $[U_{L}^{0}, U_{R}^{0}]$ that depends on the user's current observed viewpoint.
If bandwidth is not a concern, for best synthesized view quality the edge cloudlet would send to the user all camera-captured views in $\mathcal{V}$ as reference to synthesize virtual view $u$, $\forall u  \in [U_{L}^{0}, U_{R}^{0}]$. When this is not feasible due to limited bandwidth $C$ between the serving cloudlet and the user, among all subsets $\mathcal{T} \subset \mathcal{U}$ of synthesized and camera-captured views that satisfy the bandwidth constraint, the cloudlet must select the best subset $\mathcal{T}^*$ that minimizes the aggregate distortion $\mathcal{D}(\mathcal{T})$ of all virtual views $u \in [U_{L}^{0}, U_{R}^{0}]$, \textit{i.e.},
 
\begin{align}\label{eq:optimization}
{{\mathcal{T}}}^{\star} :  &\arg \min_{{\mathcal{T}}}  \mathcal{D}({\mathcal{T}}) \\
\nonumber
 &\text{s.t }  | {{\mathcal{T}}} |\leq C \\ 
\nonumber
& \hspace{5mm}{{\mathcal{T}}}\subseteq \mathcal{U} 
\end{align}
We note that (\ref{eq:optimization}) differs from existing reference view selection formulations \cite{Maugey:C14, Abreu:A15 , ren15} in that the cloudlet has the extra degree of freedom to synthesize novel virtual view(s) as new reference(s) for transmission to the user. 

Denote by $D(v)$ the distortion of viewpoint image $v$, due to lossy compression for a camera-captured view, or by DIBR synthesis for a virtual view. The distortion $\mathcal{D}(\mathcal{T})$  experienced over the navigation window  at the user is then given by 
\begin{align}\label{eq:distortion_user}
\mathcal{D}({\mathcal{T}}) = \sum_{u\in[U_{L}^{0}, U_{R}^{0}]}  \min_{v_L, v_R \in {\mathcal{T}}} \left\{ d_u(v_L,v_R, D(v_L),D(v_R)) \right\} 
\end{align}
where $D(v_L)$ and $D(v_R)$ are the respective distortions of the left and right reference views and $d_u(v_L,v_R, D_L,D_R)$ is the distortion of the virtual view $u$ synthesized using left and right reference views $v_L$ and $v_R$ with distortions $D_L$ and $D_R$, respectively. In (\ref{eq:distortion_user}), for each virtual view $u$ the best reference pair in ${\mathcal{T}}$ is selected for synthesis. Note that, unlike \cite{ren15}, the best reference pair may not be the closest references, since the quality of synthesized $u$ depends not only on the view distance between the synthesized and reference views, but also on the distortions of the references. 

\subsection{Distortion of virtual viewpoints}


We consider first an assumption on the synthesized view distortion $d_u(\,)$ called the \textit{shared optimality of reference views}:
\begin{align}\label{eq:shared_optimality}
&\text{if } d_u(v_L,v_R, D(v_L),D(v_R)) \leq d_u(v_L^{\prime},v_R^{\prime}, D(v_L^{\prime}),D(v_R^{\prime}))  
  \\  \nonumber
&\text{then } d_{u'}(v_L,v_R, D(v_L),D(v_R)) \leq d_{u'}(v_L^{\prime},v_R^{\prime}, D(v_L^{\prime}),D(v_R^{\prime}))  
\end{align}
for $\max\{v_L,v_L^{\prime}\} \leq u, u' \leq \min\{v_R,v_R^{\prime}\}$. In words, this assumption (\ref{eq:shared_optimality}) states that if the virtual view $u$ is better synthesized using the reference pair $(v_L,v_R)$ than $(v_L^{\prime},v_R^{\prime})$, then another virtual view $u^{\prime}$ is also better synthesized using $(v_L,v_R)$ than $(v_L^{\prime},v_R^{\prime})$.  

We see intuitively that this assumption is reasonable for smooth 3D scenes; a virtual view $u$ tends to be similar to its neighbor $u'$, so a good reference pair $(v_L, v_R)$ for $u$ should also be good for $u'$. We can also argue for the plausibility of this assumption as a consequence of two functional trends in the synthesized view distortion $d_v(\,)$ that are observed empirically to be generally true. For simplicity, consider for now the case where the reference views $v_L, v_R, v_L', v_R'$ have zero distortion, \textit{i.e.} $D(v_L) = D(v_R) = D(v_L') = D(v_R') = 0$. The first trend is the \textit{monotonicity in predictor's distance} \cite{cheung11tip2}; \textit{i.e.,} the further-away are the reference views to the target synthesized view, the worse is the resulting synthesized view distortion. This trend has been successively exploited for efficient bit allocation algorithms \cite{liu05, cheung11tip2}.
In our scenario, this trend implies that reference pair $(v_L, v_R)$ is better than $(v_L', v_R')$ at synthesizing view $u$ because the pair is closer to $u$, \textit{i.e.} 
\begin{align}
|u - v_L| + |v_R - u| \leq |u - v_L'| + |v_R' - u|
\label{eq:mono_refDist}
\end{align}
where $\max \{v_L, v_L' \} < u < \min \{v_R, v_R' \}$. 

It is easy to see that if reference pair $(v_L, v_R)$ is closer to $u$ than $(v_L', v_R')$, it is also closer to $u'$, thus better at synthesizing $u'$. Without loss of generality, we write new virtual view $u'$ as $u' = u + \delta$. We can then write:
\begin{align}
|(u + \delta) - v_L| + |v_R - (u + \delta)| & =  u - v_L + v_R - u \nonumber \\
& \leq u - v_L' + v_R' - u \nonumber \\
& \leq |(u + \delta) - v_L'| + |v_R' - (u + \delta)|
\end{align}
where $\max \{v_L, v_L' \} < u' < \min \{v_R, v_R'\}$.  

Consider now the case where the reference views $v_L, v_R, v_L', v_R'$ have non-zero distortions. In \cite{toni15}, another functional trend is empirically demonstrated, where a reference view $v_L$  with distortion $D(v_L)$ was well approximated as a further-away \textit{equivalent reference view} $v_L^{\#} < v_L$ with no distortion $D(v_L^{\#}) = 0$. Thus a better reference pair $(v_L, v_R)$ than $(v_L', v_R')$ at synthesizing $u$ just means that the equivalent reference pair for $(v_L, v_R)$ are closer to $u$ than the equivalent reference pair for $(v_L', v_R')$. Using the same previous argument, we see that the equivalent reference pair for $(v_L, v_R)$ are also closer to $u'$ than $(v_L', v_R')$, resulting in a smaller synthesized distortion. Hence, we can conclude that the assumption of shared optimality of reference views is a consequence of these two functional trends. 

 
  \begin{figure}[t]
 \centering{
 \subfigure[]{
 \includegraphics[width=0.42\linewidth ,draft=false]{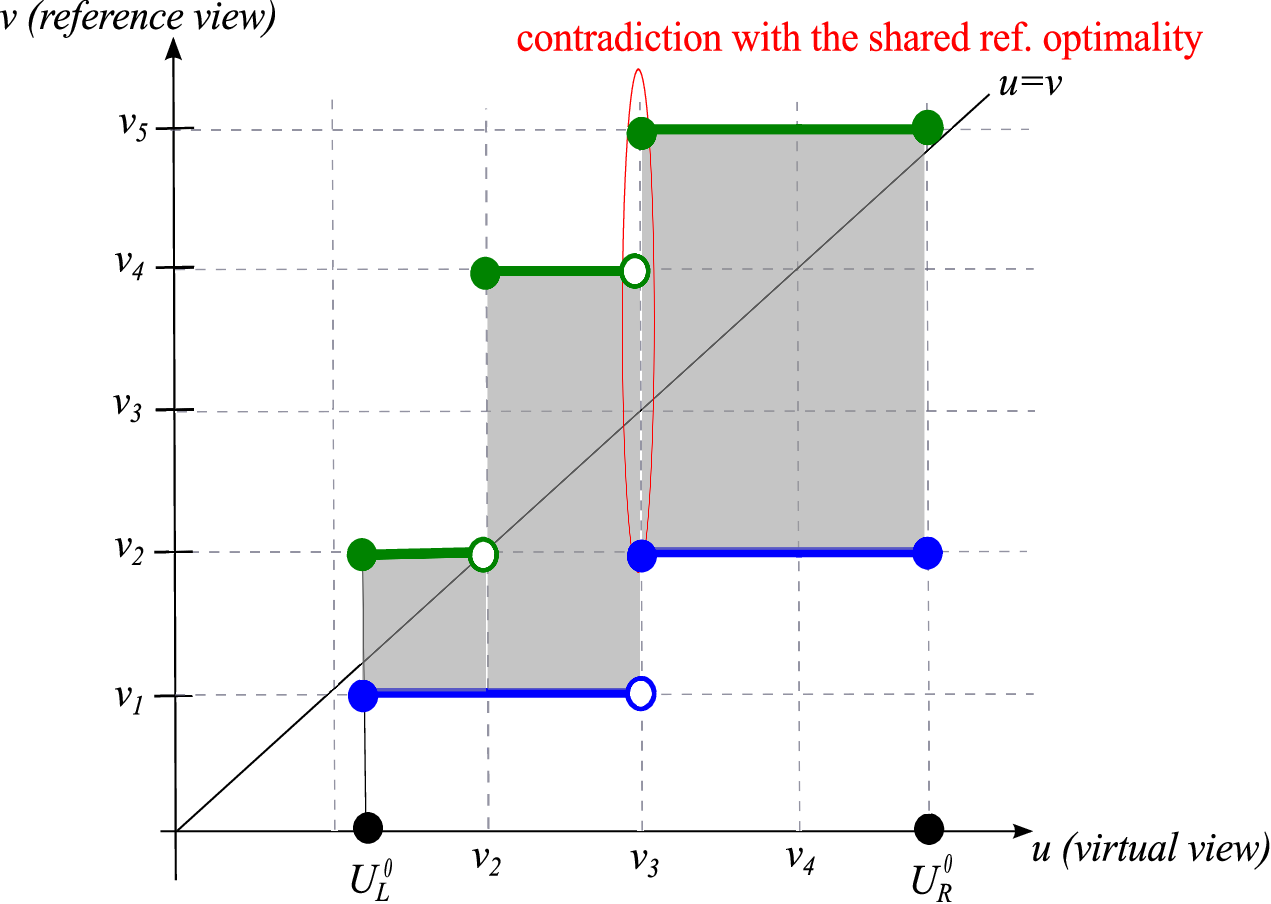} 
 \label{fig:Distorted_A}
 }
 \hfil
 \subfigure[]{
 \includegraphics[width=0.42\linewidth ,draft=false]{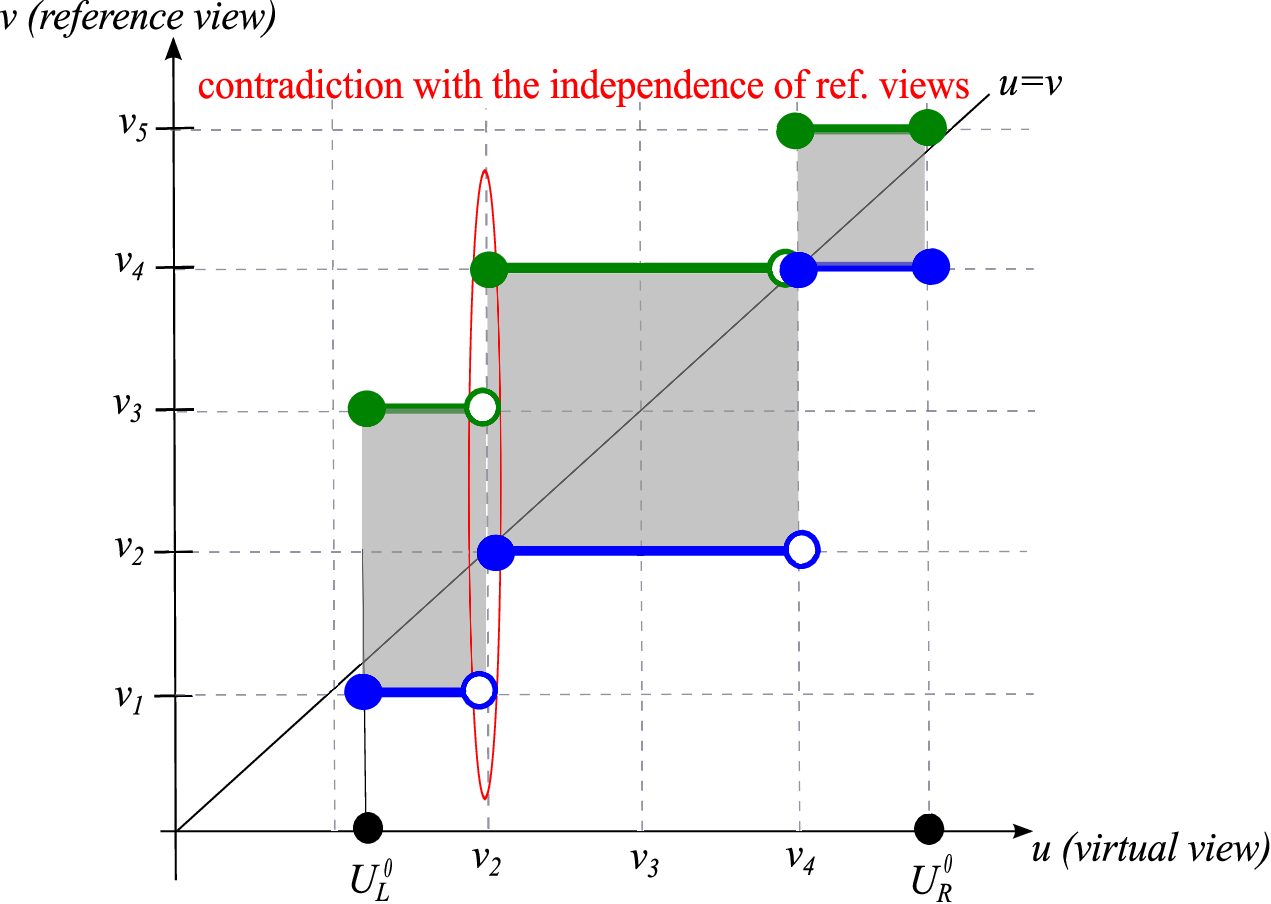} 
 \label{fig:Distorted_B}
 }}
 \hfil \caption{ Reference view assignment in (a) contradicts the shared reference assumption. Reference view assignment in (b) respects  the shared reference assumption but   contradicts the independence of reference optimality assumption.   }\label{fig:Distorted}
 \end{figure}

We can graphically illustrate possible solutions to the optimization problem \eqref{eq:optimization} under the assumption of shared optimality of reference views. Fig.\;\ref{fig:Distorted_A} depicts    the selected reference views for virtual views in the navigation window.  In the figure, the $x$-axis represents the virtual views in the window $[U_L^0, U_R^0]$ that require synthesis. Correspondingly, on the $y$-axis are two piecewise constant (PWC) functions representing the left and right reference views selected for synthesis of each virtual view $u$ in the window, assuming that for each  $u \in [U_L^0, U_R^0]$  there must be one selected reference pair $(v_L, v_R)$ such that $v_L \leq u \leq v_R$. A constant line segment---\textit{e.g.}, $v=v_1$ for $U^0_L \leq u \leq v_3$ in Fig.\;\ref{fig:Distorted_A}---means that  the same reference is used for a range of virtual views. This graphical representation results in two PWC functions---left and right reference views---above and below the $u=v$ line. The set of selected reference views are the unions of the constant step locations in the two PWC functions. 


Under the assumption of shared reference optimality we see that the selected reference views in Fig.\;\ref{fig:Distorted_A} cannot be an optimal solution. Specifically, virtual views $v_3 - 1/L$ and $v_3$ employ references $[v_1, v_4]$ and $[v_2, v_5]$ respectively.  However, if references $[v_1, v_4]$ are better than $[v_2, v_5]$ for virtual view $v_3 - 1/L$, they should be better for virtual view $v_3$ also according to shared reference optimality in \eqref{eq:shared_optimality}. An example of an optimal solution candidate under the assumption of shared reference optimality is shown in Fig.\;\ref{fig:Distorted_B}.


 \subsection{NP-hard Proof}
 \label{eq:np_hard_proof}

 We now outline a proof-by-construction that shows the reference view selection problem (\ref{eq:optimization}) is NP-hard under the shared optimality assumption. We show it by reducing the known NP-hard \textit{set cover} (SC) problem~\cite{cormen2001} to a special case of the reference view selection problem. In SC, a set of items $\mathcal{S}$ (called the universe) are given, together with a defined collection $\mathcal{C}$ of subsets of items  in $\mathcal{S}$. The SC problem is to identify at most $K$ subsets from collection $\mathcal{C}$ that covers $\mathcal{S}$, \textit{i.e.}, a smaller collection $\mathcal{C}^{\prime}\subseteq  \mathcal{C} $ with $|\mathcal{C}^{\prime}| \leq K$ such that every item in $\mathcal{S}$ belongs to at least one subset in collection $\mathcal{C}^{\prime}$. 

\begin{figure}[t]
 \begin{center}
 \includegraphics[width=0.4\linewidth ,draft=false]{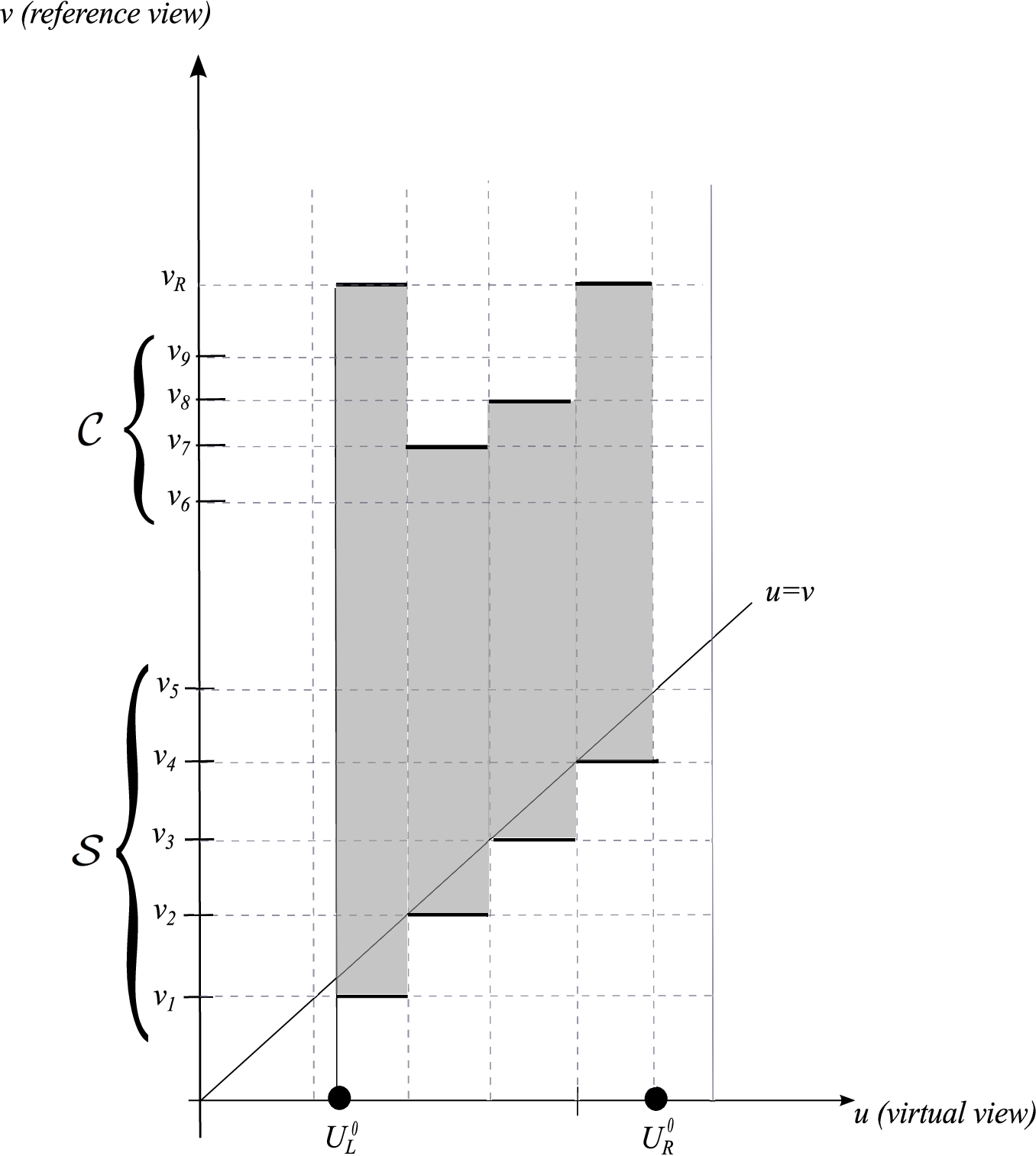} 
 \caption{Example of items set $\mathcal{S}$ and collection of sets  $\mathcal{C}$, with $|\mathcal{S}|= 5$ and $|\mathcal{C}|= 4$.} \label{fig:NP_Hardness}
 \end{center}
  \end{figure}

We construct a corresponding special case of our reference view selection problem as follows. For each item $i$ in $\mathcal{S} = \{1, \ldots, |\mathcal{S}|\}$ in the SC problem, we first construct an undistorted reference view $i$. In addition, we construct a default undistorted right reference view $|\mathcal{S}|+1$, and the navigation window is set to $[U^0_L, U^0_R] = [1, |\mathcal{S}|+1]$ and $L=2$. Further, for each item $i$ in $\mathcal{S}$, we construct a virtual view $i + \frac{1}{2}$ that requires the selection of left reference $i$, in combination of default right reference $|\mathcal{S}|+1$, for the resulting synthesized view distortion $d_{i + \frac{1}{2}}(i,|\mathcal{S}|+1,0,0)$ to achieve distortion $\bar{D} < \infty$. Thus the selection of $|\mathcal{S}|$ left references and one default right reference $|\mathcal{S}|+1$ consumes $|\mathcal{S}|+1$ views worth of bandwidth already. See Fig.~\ref{fig:NP_Hardness} for an illustration. Note that given this selection of left reference views, any selection of right reference views will satisfy the shared optimality of reference views assumption. 

For each subset $j$ in collection $\mathcal{C} = \{1, \ldots, |\mathcal{C}|\}$ in the SC problem, we construct a right reference view $|\mathcal{S}| + 1 + j$, such that if item $i$ belongs to subset $j$ in the SC problem, the synthesized distortion $d_{i + \frac{1}{2}}(i,|\mathcal{S}|+1+j,0,0)$ at virtual view $1 + \frac{1}{2}$ will be reduced to $\bar{D} - \Delta$ given right reference view $|\mathcal{S}| + 1 + j$ is used. The corresponding binary decision we ask is: given channel bandwidth of $|\mathcal{S}| + 1 + K$, is there a reference view selection such that the resulting synthesized view distortion is $|\mathcal{S}| (\bar{D} - \Delta)$ or less?

From construction, it is clear that to minimize overall distortion, left reference views $1, \ldots, |\mathcal{S}|$ and default right reference view $|\mathcal{S}| + 1$ must be first selected in any solution with distortion $< \infty$. Given remaining budget of $K$ additional views, if distortion of $|\mathcal{S}| (\bar{D} - \Delta)$ is achieved, that means $K$ or fewer additional right reference views are selected to reduce synthesized distortion from $\bar{D}$ to $\bar{D} - \Delta$ at each of the virtual view $i + \frac{1}{2}$, $i \in \{1, \ldots, |\mathcal{S}|\}$. Thus these additionally $K$ or fewer selected right reference views correspond exactly to the subsets in the SC problem that covers all items in the set $\mathcal{S}$. This solving this special case of the reference view selection problem is no easier than solving the SC problem, and therefore the reference view selection problem is also NP-hard. $\Box$

%% file: algo_approx.tex
Given that the reference view selection problem (\ref{eq:optimization}) is NP-hard under the assumption of shared optimality of reference views, in this section we introduce another assumption on the synthesized view distortion that holds in most common 3D scenes. Given these two assumptions, we show that (\ref{eq:optimization}) can now be solved optimally in polynomial time by a \ac{DP} algorithm. We also analyze the DP algorithm's computation complexity.

\subsection{Independence of reference optimality assumption}

The second assumption on the synthesized view distortion $d_u(\,)$ is the \textit{independence of reference optimality}, stated formally as follows:
\begin{align}\label{eq:indep_optimality}
&\text{if } d_u(v_L,v_R, D(v_L),D(v_R)) \leq d_u(v_L^{\prime},v_R, D(v_L^{\prime}),D(v_R))  
  \\  \nonumber
&\text{then } d_u(v_L,v_R^{\prime}, D(v_L),D(v_R^{\prime})) \leq d_u (v_L^{\prime},v_R^{\prime}, D(v_L^{\prime}),D(v_R^{\prime})) 
\end{align}
for $\max\{v_L,v_L^{\prime}\} \leq u  \leq \min\{v_R,v_R^{\prime}\}$.
In words, the assumption (\ref{eq:indep_optimality}) states that if $v_L$ is a better left reference than $v_L'$ when synthesizing virtual view $u$ using $v_R$ as right reference, then $v_L$ remains the better left reference to synthesize $u$ even if a different right reference $v_R'$ is used. This assumption essentially states that contributions towards the synthesized image from the two references are independent from each other, which is reasonable since each rendered pixel in the synthesized view is typically copied from one of the two references, but not both.
\ifimportant
\else
We can also argue for the plausibility of this assumption as a consequence of the two aforementioned functional trends in the synthesized view distortion $d_v(\,)$ in Section \ref{sec:algo}. Consider first the case where the reference views $v_L, v_R, v_L', v_R'$ have zero distortion. The \textit{monotonicity in predictor's distance} in \eqref{eq:mono_refDist} for a common right reference view becomes
\begin{align}\label{eq:dist_assumpt2}
|u - v_L| + |v_R - u| &\leq |u - v_L'| + |v_R - u|  \hspace{5mm} \longrightarrow  \hspace{5mm} |u - v_L|    \leq |u - v_L'|   
\end{align}
where $\max \{v_L, v_L' \} < u <   v_R$. Thus if $v_L$ is preferred to $v_L'$ for  $v_R>u$, it will hold also for  $v_R'$ as long as $v_R'>u$. 
Consider now the case where the reference views $v_L, v_R, v_L', v_R'$ have non-zero distortions. Introducing  the \textit{equivalent reference views} $v_L^{\#} < v_L$    with no distortion $D(v_L^{\#}) = 0$, the same argument of  \eqref{eq:dist_assumpt2} holds for the equivalent reference views, leading to $|u - v_L^{\#}|    \leq |u - v_L^{\prime\#}|$, $\forall v_R>u$. 
\fi

 \begin{figure}[t]
 \centering{
 \subfigure[]{
 \includegraphics[width=0.42\linewidth ,draft=false]{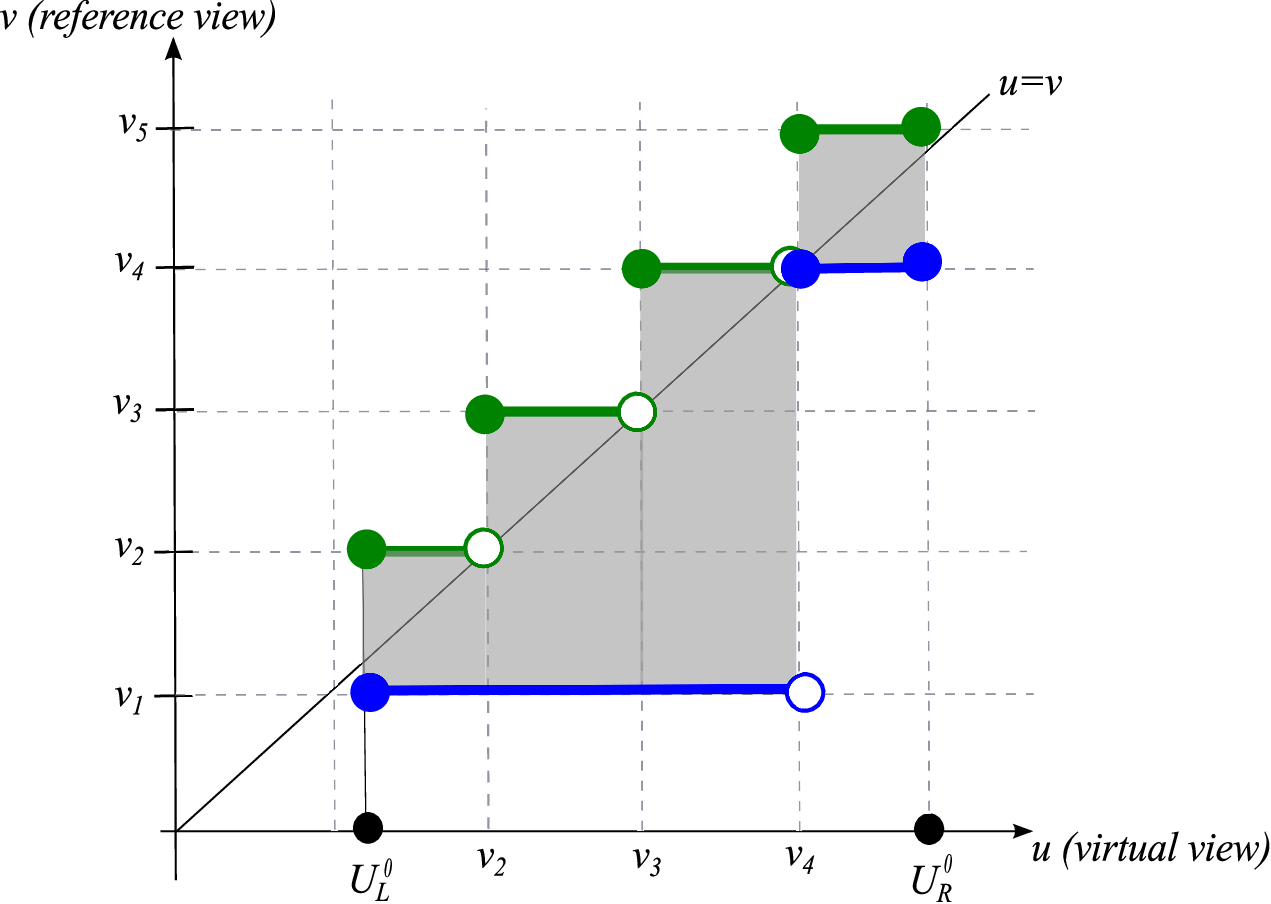} 
 \label{fig:Distorted_C}
 }
 \hfil
 \subfigure[]{
 \includegraphics[width=0.42\linewidth ,draft=false]{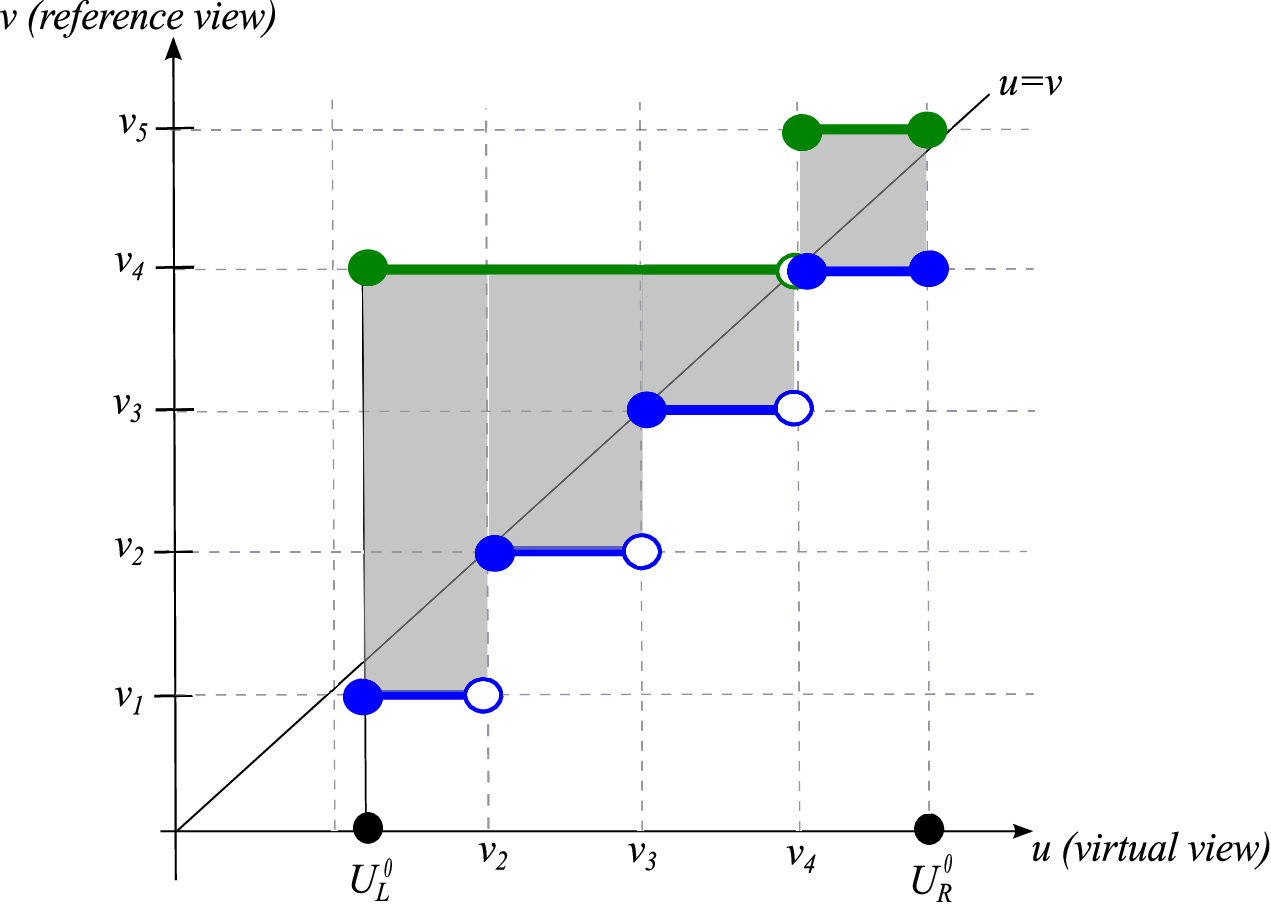} 
 \label{fig:Distorted_D}
 }}
 \hfil \caption{
  Reference view assignments in (a) and (b) are optimal solution candidates under both assumptions. We name these two cases ``shared-left" and ``shared-right", respectively.}\label{fig:Distorted_Region}
 \end{figure}

We illustrate different optimal solution candidates to \eqref{eq:optimization} now under both virtual view distortion assumptions to impart intuition. 
We see that the assumption of independence of reference optimality would prevent the reference view selection in Fig.\;\ref{fig:Distorted_B} from being an optimal solution. Specifically, we see that both $v_3$ and $v_4$ are feasible right reference views for virtual views $v_2 - 1/L$ and $v_2$. Regardless of which left references are selected for these two virtual views, if $v_3$ is a strictly better right reference than $v_4$, then having both virtual views select $v_3$ as right reference will result in a lower overall distortion (and vice versa). If $v_3$ and $v_4$ are \textit{equally} good right reference views resulting in the \textit{same} synthesized view distortion, then selecting just $v_4$ without $v_3$ can achieve the same distortion with one fewer right reference view. Thus the selected reference views in Fig.\;\ref{fig:Distorted_B} cannot be optimal.

We can thus make the following observation: as virtual view $u$ increases, an optimal solution cannot switch right reference view from current $v_R$ earlier than $u = v_R$. Conversely, as virtual view $u$ decreases, an optimal solution cannot switch left reference view from current $v_L$ earlier than $u = v_L - 1/L$. As examples, Fig.\;\ref{fig:Distorted_Region}   provides  solutions of left and right reference views for virtual views in the navigation window.  In the figure, on the $x$-axis are the virtual views $u$ in the window $[U_L^0, U_R^0]$ that require synthesis. Correspondingly, on the $y$-axis are the left and right reference views (blue and red PWC functions respectively) selected to synthesize each virtual view $u$ in the window. 
We see that the reference view selections in Fig.\;\ref{fig:Distorted_C} and Fig.\;\ref{fig:Distorted_D} are optimal solution candidates to \eqref{eq:optimization}. Thus, the optimal reference view selections must be graphically composed of ``staircase" virtual view ranges as shown in Fig.\,\ref{fig:Distorted_C} and Fig.\,\ref{fig:Distorted_D}. In other words, either a shared left reference view $v^s_L$ is used for multiple virtual view ranges $[u_i, u_{i+1})$ where each range has the same $v^s_L$ as left reference (``shared-left" case), or a shared right reference view $v^s_R$ is used for multiple ranges $[u_i, u_{i+1})$, where each range has $v^s_R$ as its right reference (``shared-right" case). This motivates us to design an efficient DP algorithm to solve \eqref{eq:optimization} optimally in polynomial time.

\subsection{DP Algorithm}

We first define a recursive function $\Phi(u_L, v_L, k)$ as the minimum aggregate synthesized view distortion of views between $u_L$ and $U_R^0$, given $v_L$ is the selected left reference view for synthesizing view $u_L$, and  there is a budget of $k$ additional reference views. To analyse $\Phi(u_L, v_L, k)$, we consider the two ``staircase" cases identified by Fig.\,\ref{fig:Distorted_C} and Fig.\,\ref{fig:Distorted_D} separately, and show how $\Phi(u_L, v_L, k)$ can be evaluated in each of the cases.

Consider first the ``shared-left" case (Fig.\,\ref{fig:Distorted_C}) where a shared left reference view is employed in a sequence of virtual view ranges.  A view range represents a contiguous range of virtual viewpoints that employ the same left and right reference views.    The algorithm selects a new right reference view $v$, $v > u_L$, creating a new range of virtual views $[u_L, v)$. Virtual views in range $[u_L, v)$ are synthesized using a shared left reference $v_L$ and the newly selected reference view $v$, resulting in distortion $d_u(v_L, v, D(v_L), D(v))$ for each virtual view $u$, $u_L \leq u < v$.  The aggregate distortion function $\Phi(u_L, v_L, k)$ for this case is the distortion of views in  $[u_L, v)$ plus a recursive term $\Phi(v, \Lambda(v_L, v), k-1)$ to account for aggregate  synthesized view distortions to the right of $v$:
\begin{equation}\label{eq:recurs_1}
\sum_{u=u_L}^{v - \frac{1}{L}} d_u(v_L, v, D(v_L), D(v)) + \Phi(v, \Lambda(v_L, v), k-1)
\end{equation}
where $k-1$ is  the remaining budget of additional reference views, and  $\Lambda(v_1, v_2)$ chooses the better of the two left reference views, $v_1$ and $v_2$, for the recursive function $\Phi(\,)$. In particular, using any right reference view $v_R$ and virtual view $u$, where $\max \{v_1, v_2\} < u < v_R$, we set $\Lambda(v_1, v_2) = v_1$ if virtual view $u$ is better synthesized using  $v_1$ as left reference than $v_2$ (and set $\Lambda(v_1, v_2) = v_2$ otherwise). Formally, the left reference selection function $\Lambda(v_1, v_2)$ is defined as:
\begin{equation}
\Lambda(v_1, v_2) = \left\{ \begin{array}{ll}
v_1 & \mbox{if} ~~ d_u(v_1, v_R, D(v_l), D(v_R)) \leq d_u(v_2, v_R, D(v_2), D(v_R)) \\
v_2 & \mbox{o.w.}
\end{array} \right.
\end{equation}
Given our two assumptions, we know that the selected left reference $\Lambda(v_1, v_2)$ remains better for all other virtual views $u$ in $[\max \{v_1, v_2\}, v_R]$.

We now consider the ``shared-right" case  (Fig.\,\ref{fig:Distorted_D}) where a newly selected view $v$ is actually a common right reference view for a sequence of virtual view ranges from $u_L$ to $v$. We first define a companion recursive function $\Psi(u_L, v_L, v_R, n)$ that returns the minimum aggregate synthesized view distortion from view $u_L$ to $v_R$, given that $v_L$ is the selected left reference view, $v_R$ is the common right reference view, and there is a budget of $n$ other left reference views in addition to  $v_L$. We can write $\Psi(u_L, v_L, v_R, n)$ recursively as follows:
\begin{equation} 
\Psi(u_L, v_L, v_R, n) = \left\{ 
\begin{array}{ll}
\min\limits_{v > u_L}
\sum\limits_{u = v_L}^{v - \frac{1}{L}} d_u(v_L, v_R, D(v_L), D(v_R)) +
\Psi(v, v, v_R, n-1)
& \mbox{if} ~~ k \geq 1 \\
\sum\limits_{u = v_L}^{v_R - \frac{1}{L}} d_u(v_L, v_R, D(v_L), D(v_R)) 
& \mbox{o.w.}
\end{array}
\right.
\label{eq:Psi}
\end{equation}
In more details, the equation  (\ref{eq:Psi}) states that $\Psi(u_L, v_L, v_R, n)$ is the synthesized view distortion  of views in the range $[u_L, v)$, plus the recursive distortion $\Psi(v, v, v_R, n-1)$ from view $v$ to $v_R$ with a reduced reference view budget $n-1$. 

We can now put the two cases together into a complete definition of $\Phi(u_L, v_L, k)$ as follows:
\begin{align}\label{eq:optimal_2}
\Phi(u_L, v_L, k) & = \min_{v > v_L} \left\{ \min \left[ \underbrace{ 
\sum_{u = u_L}^{v - \frac{1}{L}} d_u(v_L,v, D(v_L), D(v)) + \Phi(v, \Lambda(v_L,v), k-1)}_{\text{``shared-left" case }}, 
\right.
\right. \\ \nonumber
&\hspace{2cm} \left.  \left. 
\underbrace{\min_{1 \leq n\leq k-1}  
\Psi(u_L, v_L, v, n) + \Phi(v, v, k-n-1)}_{\text{``shared-right" case }}
\right]
\right\}
\end{align}
The relation  \eqref{eq:optimal_2}  states that $\Phi(u_L, v_L, k)$ examines each candidate reference view $v$, $v > v_L$, which can be used either as right reference for synthesizing virtual views in $[u_L, v)$ with left reference $v_L$ (``shared-left" case), or as a common right reference for a sequence of $n+1$ virtual view ranges within the interval $[u_L, v)$ (``shared-right" case).

When the remaining view budget is $k=1$, the relation in  \eqref{eq:optimal_2} $\Phi(u_L, v_L, 1)$ simply selects a right reference view $v$, $v \geq U_R^0$, which minimizes the aggregate synthesized view distortion for the range $[u_L, U_R^0]$:
\begin{equation}
\Phi(u_L, v_L, 1) = \min_{v \geq U_R^0}
\sum_{u=u_L}^{U_R^0} d_u(v_L, v, D(v_L), D(v))
\end{equation}

Having defined $\Phi(u_L, v_L, k)$, we can identify the best $K$ reference views  by calling $\Phi(U_L^0, v, K)$ repeatedly to identify the best leftmost reference view $v$, $v \leq U_L^0$, and start the
selection of the $K-1$ remaining  reference views as follows 
\begin{equation}
\min_{v \leq U_L^0} \Phi(U_L^0, v, K-1)
\end{equation}


\subsection{Computation Complexity}

Our proposed DP algorithm requires two different tables to be stored.  The first time $\Psi(u_L, v_L, v_R, n)$ is computed, the result can be stored in entry $[(u_L-U_L^0)/L][(v_L-U_L^0)/L][(v_R-U_L^0)/L][n]$ of a DP table ${\boldsymbol\Psi}^*$, so that subsequent calls with the same arguments can be simply looked up. Analogously,  the first time $\Phi(u_L, v_L, k)$ is called, the computed value is stored in entry $[(u_L-U_L^0)/L][(v_L-U_L^0)/L][k]$ of another DP table ${\boldsymbol\Phi}^*$ to avoid repeated computation in future recursive calls.

We bound the computation complexity of our proposed algorithm (\ref{eq:optimal_2}) by computing a bound on the sizes of the required DP tables and the cost in computing each table entry. For notation convenience, let the number of reference views and synthesized views be $S_v = (V-1)/L$ and $S_u = (U_R^0 - U_L^0)/L$, respectively. The size of DP table ${\boldsymbol\Phi}^*$ is no larger than $S_u \times S_v \times K$. The cost of computing an entry in ${\boldsymbol\Phi}^*$ using (\ref{eq:optimal_2}) over all possible reference views $v$ involves the computation of the  ``shared-left" case with complexity $O(S_u)$ and the one of the ``shared-right" case  with complexity $O(K)$. Thus, each table entry has complexity $O(S_v S_u + S_v K)$. Hence the complexity of completing the DP table ${\boldsymbol\Phi}^*$ is $O(S_u^2 S_v^2 K + S_u S_v ^2 K^2)$. Given that in typical setting $S_u \gg K$, the complexity for computing DT table ${\boldsymbol\Phi}^*$ is thus $O(S_u^2 S_v^2 K)$.

We can perform similar procedure to estimate the complexity in computing DP table ${\boldsymbol\Psi}^*$. The size of the table in this case is upper-bounded by $S_u \times S_v \times S_v \times K$. The complexity in computing each entry is $O(S_u)$. Thus the complexity of computing DP table ${\boldsymbol\Psi}^*$ is $O(S_u^2 S_v^2 K)$. which is the same as DP table ${\boldsymbol\Phi}^*$. Thus the overall computation complexity of our solution in  (\ref{eq:optimal_2}) is also $O(S_u^2 S_v^2 K)$. 

%% file: results.tex
\begin{table*}[t]
\begin{center}
\caption{Viewpoints notation. } \label{table:views_notation}
\begin{tabular}{|c||c|c|c|c|c|c|c|c|c|c|c|c|} \hline 
Camera ID as in \cite{Kim:SRLF:2013}, ``Statue" & $ {50}$ & $51$  & $52$ & $53$ & $54$ & $55$ & $56$ & $57$ & $ {58}$ & $59$ & $\ldots$ & $ {98}$      \\ \hline
Camera ID as in \cite{Kim:SRLF:2013}, ``Mansion" & $ {25}$ & $26$  & $27$ & $28$ & $29$ & $30$ & $31$ & $32$ & $ {33}$ & $34$ & $\ldots$ & $ {73}$      \\ \hline\hline
Camera ID in our work & $ {0}$ & $1.125$  & $1.25$ & $1.375$ & $1.5$ & $1.625$ & $1.75$ & $1.875$ & $ {1}$ & $2.125$ & $\ldots$ & $ {6}$    \\ \hline 
\end{tabular}
\end{center}
\end{table*}
 
\subsection{Settings}
We study the performance of our algorithm and we show the distortion gains offered by cloudlets-based virtual   view  synthesis. For a given   navigation window $[U_L^0,U_R^0]$, we provide the average quality   at which   viewpoints in the navigation window is synthesized.  This means that we evaluate the average distortion of the navigation window   as $(1/N)\sum_{u=U_L^0}^{U_R^0} d_u$, with $N$ being the number of synthesized viewpoints in the  
navigation window, and we then compute the corresponding PSNR.
In our algorithm, we have considered the following  model for the distortion of the synthesized viewpoint $u$ from   reference views $V_L$, $V_R$  
\begin{align}
\label{eq:distortion_model}
 d_u (V_L, V_R, D_L, D_R)  = \alpha D_{min} + (1-\alpha) \beta D_{max} + 
\left[ 1-  \alpha  -  (1-\alpha) \beta \right] D_{I}
\end{align}
where $D_{min}=\min\{D_L,D_R\}$, $D_{max}=\max\{D_L,D_R\}$, $D_I$ is the inpainted distortion, and 
$ 
 \alpha = \exp\left(-\gamma |u-V_{min}| d\right), \,
 \beta = \exp\left(-\gamma |u-V_{max}| d\right) \nonumber
$ 
with  $d$ is the distance between two consecutive camera views $v_i$ and $v_{i}+1$,    $V_{min}=V_L$ if $D_L\leq D_R$, $V_{min}=V_R$ otherwise, and $V_{max}=V_L$ if $D_L>D_R$, $V_{max}=V_R$. 
The model can be explained as follows. A virtual synthesis $u$, when reconstructed from $(V_L,V_R)$ has a relative portion $\alpha\in[0,1]$ that is reconstructed  at a distortion $D_{min}$, from the dominant reference view, defined as the one with minimum distortion. The remaining portion of the image, i.e., $1-\alpha$, is either reconstructed by the non-dominant reference view for a potion  $\beta$, at a distortion $D_{max}$, or it is inpainted, at a  distortion $D_I$.

The   results have been carried out using   3D sequences   ``Statue" and ``Mansion"~\cite{Kim:SRLF:2013},  where $51$ cameras acquire the scene with uniform spacing between the camera positions. The spacing between camera positions is $5.33$ mm and $10$ mm for   ``Statue" and ``Mansion", respectively. Among all camera views provided for both sequences, only a subset represents the set of camera views $\mathcal{V}$ available at the cloudlet,   while the remaining are virtual views to be synthesized.   Table \ref{table:views_notation} depicts how the camera  notation used in \cite{Kim:SRLF:2013}   is adapted to our notation.  Finally, for the ``Mansion" sequence, in the theoretical model in \eqref{eq:distortion_model} we used $\beta=0.2$, $D_{max}=450$, and $d=50$, while for the ``Statue" sequence we used  $\beta=0.2$, $D_{max}=100$, and $d=25$.

 In the following, we compare the performance achieved by   virtual view synthesis   in the  cloudlets with respect to the scenario in which cloudlets only send to users a subset of camera views. 
We denote by $\mathcal{T}_{s}$ the subset of selected reference views when synthesis is allowed in the network, and by  $\mathcal{T}_{ns}$ the subset of selected reference views when only camera views can be sent as reference views, i.e., when synthesis is not allowed in the network.    For both the cases of network synthesis and no network  synthesis, the best subset of reference views is evaluated both with the proposed  view selection algorithm and   with an exact solution, i.e., an exhaustive search of all possible combinations of reference views. For the proposed algorithm, the distortion is evaluated both with  experimental computation of the distortion, where the  results are labeled ``Proposed Alg. (Experimental Dist)",  and with the   model in     \eqref{eq:distortion_model}, results labeled ``Proposed Alg. (Theoretical Dist)".  
For all three   algorithms, once the optimal subset of reference view is selected,  the full navigation window is reconstructed experimentally and the mean PSNR of the actual reconstructed sequence is computed.

In the following,   we first validate the   distortion model in   \eqref{eq:distortion_model} as well as   the proposed optimization algorithm. Then, we provide simulation using the model in   \eqref{eq:distortion_model} and study the gain offered by network synthesis. For the sake of clarity in the notation, in the  following we identify the viewpoints by their indexes only. This means that the set of camera views $\{v_0, v_1, v_3\}$, for example, is denoted in the following by $\{0, 1, 3\}$. Analogously for   the navigation window  $[u_{0.75}, u_{5.25} ]$ is denoted in the following by $[{0.75}, {5.25} ]$.

\subsection{Performance of the view selection algorithm}
\begin{figure}[t]
\subfigure[Mansion]
{\label{fig:Mansion_validation_equallySpaced}
\includegraphics[width=0.48\linewidth ,draft=false]{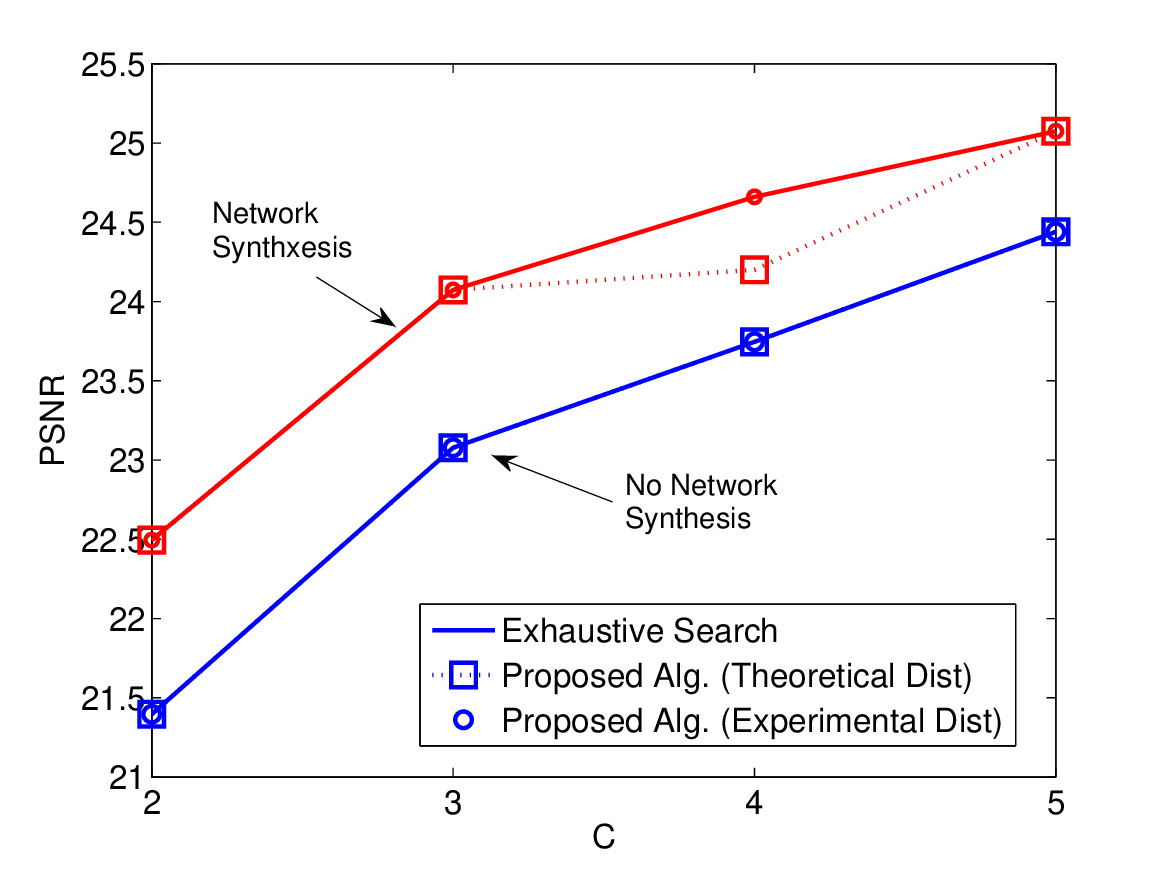}  
}  \hfill   
\subfigure[Statue]{\label{fig:Statue_validation_equallySpaced}
\includegraphics[width=0.48\linewidth ,draft=false]{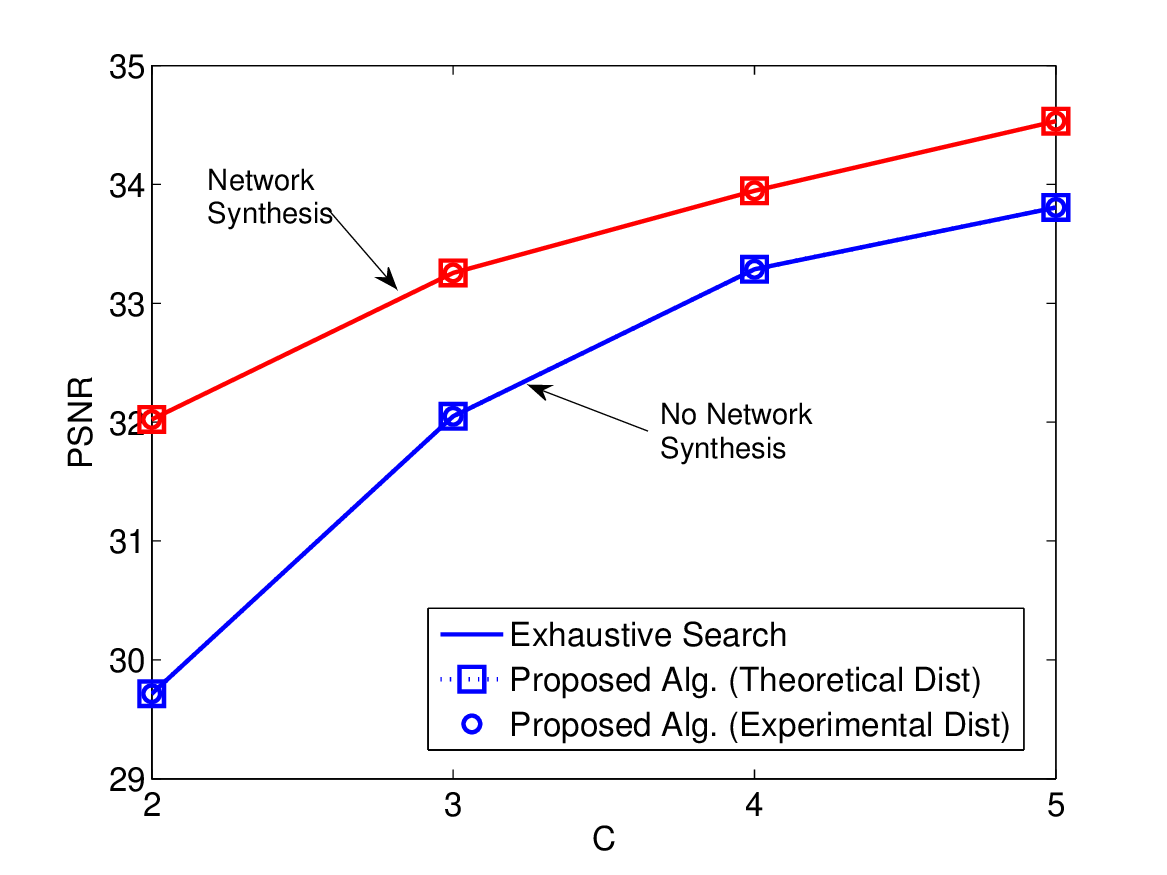}
}   \caption{Validation of the proposed optimization model  with  equally spaced cameras set  $\mathcal{V}=\{0,1, 2, \ldots, 5, 6\}$, and  a  navigation window $[0.75,5.25]$ for ``Mansion"   and  ``Statue" sequences.  } \label{fig:Eq_Spaced_Validation}
\end{figure}

In Fig. \ref{fig:Eq_Spaced_Validation}, we provide the mean PSNR as a function of the available bandwidth $C$  in the setting of  a regular spaced cameras set  $\mathcal{V}=\{0, 1, 2, \ldots, 5, 6\}$, and  a navigation window   $[{0.75}, {5.25}]$ requested by the user. Results are provided for   the ``Mansion" and the  ``Statue" sequences in   Fig. \ref{fig:Mansion_validation_equallySpaced} and Fig. \ref{fig:Statue_validation_equallySpaced}, respectively. 
For the ``Mansion" sequence, the proposed algorithm with experimental distortion perfectly matches the exhaustive search. Also the proposed algorithm based on theoretical distortion nicely matches the exhaustive search method, with the exception of the experimental point at $C=4$ in the network synthesis case. In that experiment, the algorithm selects as best subset $\mathcal{T}_s=\{ {0.75},  2,   4,  {5.25}\}$ rather than $\mathcal{T}_{s}=\{ {0.75}, 2,  3, {5.25}\}$ selected by the exhaustive search. Beyond the good match between exhaustive search and proposed algorithm, Fig. \ref{fig:Mansion_validation_equallySpaced} also shows the gain achieved in synthesizing reference views at the cloudlets. For $C=2$, the optimal sets of reference views are $\mathcal{T}_s=\{ {0.75},    {5.25}\}$ and $\mathcal{T}_{ns}=\{ 0, 6\}$. The possibility of selecting the view at position $0.75$ as reference view   reduced the reference view distance for  viewpoints in   $[ {0.75},  {5.25}]$ compared to the case in which camera view $0$ is selected. Thus, as long as the viewpoint $ {0.75}$ is synthesized at a good quality  in the network,  synthesizing in the network improves the quality of the reconstructed region of interest, when the bandwidth $C$ is limited. 
 Increasing the channel capacity reduces the quality gain between synthesis and no synthesis at the cloudlets. For $C=4$, for  example, the virtual viewpoint $0.75$ is used to reconstruct the views range $[0.75,2)$ of the navigation window. Thus, the benefit of   selecting $0.75$ rather than $0$ is limited to a portion of the navigation window and this portion usually decreases for large $C$. 
Similar considerations can be derived from Fig. \ref{fig:Statue_validation_equallySpaced},  for the ``Statue"  sequence. We observe a very good match between the proposed algorithm and the exhaustive search one.  

\begin{figure}
\begin{center}
\includegraphics[width=0.55\linewidth ,draft=false]{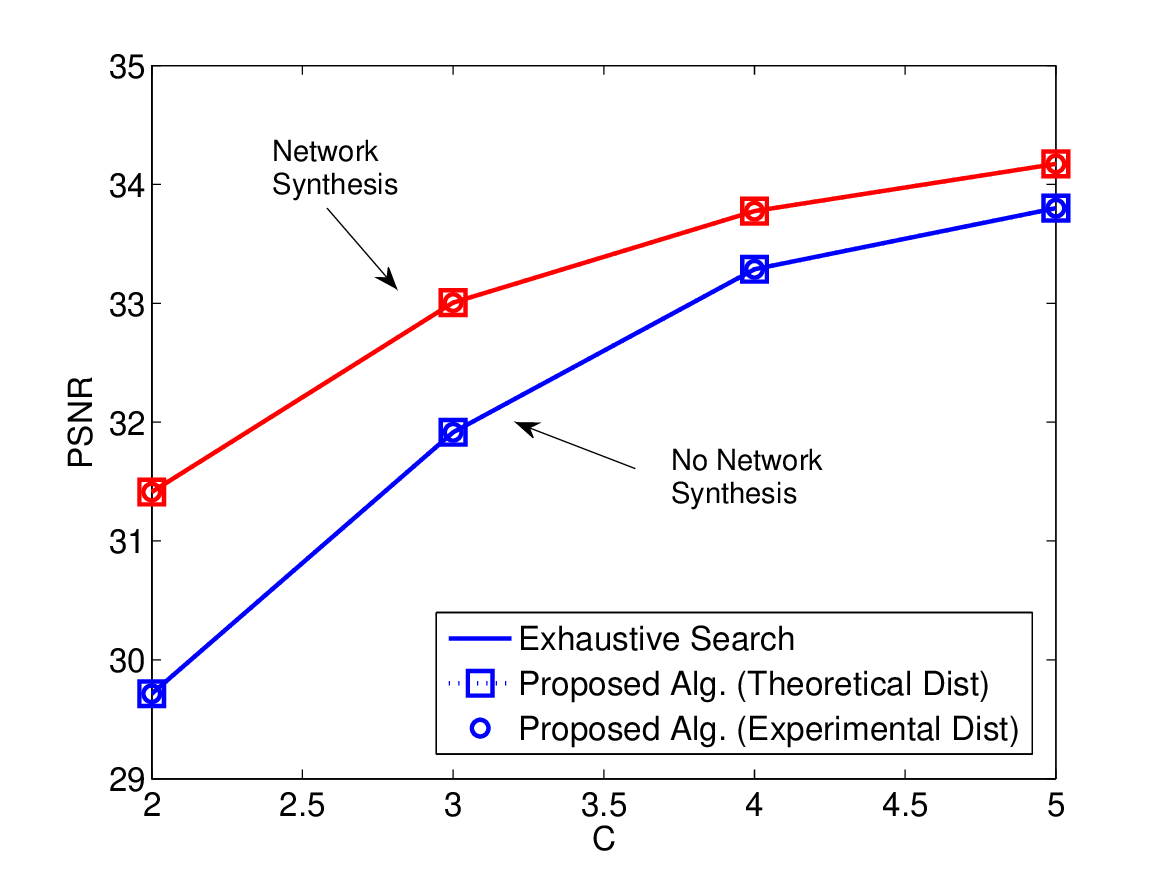}
  \caption{Validation of the proposed optimization model for ``Statue" sequence with unequally spaced cameras $\mathcal{V}=\{0,1.5, 2, 2.75, 4, 5, 6\}$ and a     navigation window  $[0.75,5.25]$. 
   } \label{fig:Statue_Validation}
\end{center}
\end{figure}

We  then compare in   Fig. \ref{fig:Statue_Validation} the performance of  the exhaustive search algorithm with our optimization method in the case of non-equally spaced cameras. The ``Statue" sequence is considered with unequally spaced cameras set  $\mathcal{V}=\{0,1.5, 2, 2.75, 4, 5, 6\}$, and   a navigation window $[0.75,5.25]$ at the client.   Similarly to the equally spaced scenario, the performance of proposed optimization  algorithm matches the one of the exhaustive search.  This confirms the validity of our assumptions and the optimality of the DP optimization solution.  Also in this case, a  quality gain is  offered by virtual view synthesis in the network, with a  maximum gain  achieved for $C=2$, with   optimal reference views $\mathcal{T}_s=\{0.75,5.25\}$ and $\mathcal{T}_{ns}=\{0,6\}$. 

\subsection{Network synthesis gain}
  Now, we aim at studying the   performance gain due to   synthesis  in the network for different   scenarios.
 However,   multiview video sequences (with both texture and depth maps) currently  available as test sequences  have a very limited number of views (e.g., $8$ views in the Ballet video sequences\footnote{http://research.microsoft.com/en-us/um/people/sbkang/3dvideodownload/}). Because of the lack of test sequences, we consider synthetic scenarios and we adopt the distortion model in   \eqref{eq:distortion_model} both for  solving the optimization algorithm and evaluating the system performance. 
The following results are   meaningful since we already validated our synthetic distortion model in the previous subsection.

\begin{figure}
\centering
\includegraphics[width=0.55\linewidth ,draft=false]{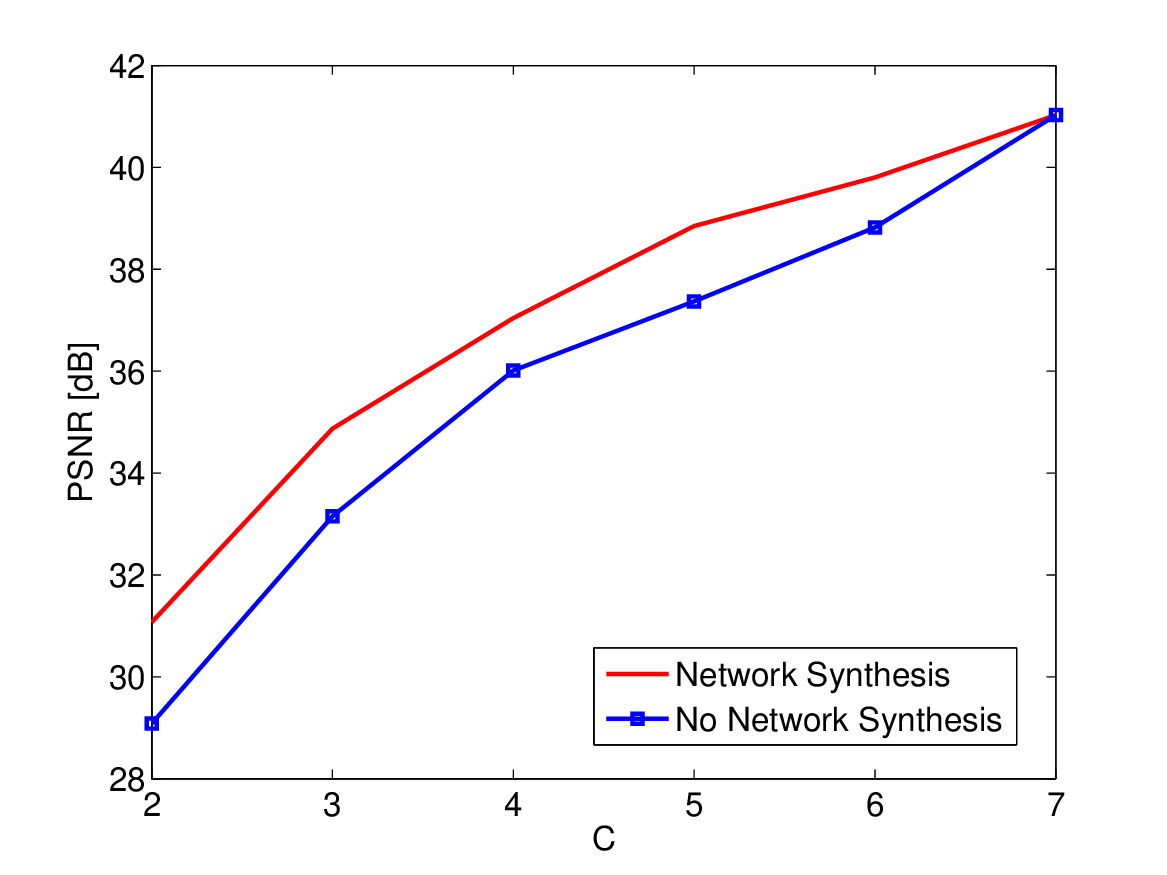} 
\caption{PSNR (in dB)  as a function of the channel capacity $C$ for different channel capacity values  $C$ for a regular spaced camera set with varying distance among cameras, $\gamma=0.3$, $D_I=300$,   navigation window $[0.75,5.25]$, and camera set $\mathcal{V}=\{0,1, 2, \ldots, 5, 6\}$ (equally spaced cameras). }\label{fig:PSNR_vs_C_varying_d}
\end{figure}

  \begin{table}[t]
 \begin{center}
\caption{Optimal subsets for the scenario of Fig. \ref{fig:PSNR_vs_C_varying_d}.}
\begin{tabular}{|c||c||c|c|c|c|c|c|c|c|c|c|} \hline 
$C $ & $\mathcal{T}_{s}$ &      $\mathcal{T}_{ns}$        \\ \hline\hline
$2$  & $\{0.75, 5.25\}$ &     $\{\pmb{0}, \pmb{6}\}$       \\ \hline
$3$  & $\{0.75, \pmb{3},  5.25\}$ &       $\{\pmb{0}, \pmb{3}, \pmb{6}\}$       \\ \hline
$4$  & $\{0.75, \pmb{2}, \pmb{4}, 5.25\}$  &     $\{\pmb{0}, \pmb{2}, \pmb{4}, \pmb{6}\}$       \\ \hline
$5$  & $\{0.75, \pmb{2}, \pmb{3}, \pmb{4}, 5.25\}$ &   $\{\pmb{0}, \pmb{2}, \pmb{3}, \pmb{4}, \pmb{6}\}$      \\ \hline
$6$  & $\{ \pmb{0},  \pmb{1},  \pmb{2}, \pmb{3}, \pmb{4}, 5.25\}$ &    $\{\pmb{0}, \pmb{1}, \pmb{2}, \pmb{3}, \pmb{4}, \pmb{6}\}$     \\ \hline
$7$  & $\{ \pmb{0},  \pmb{1}, \pmb{2}, \pmb{3}, \pmb{4}, \pmb{5}, \pmb{6}\}$ &      $\{\pmb{0}, \pmb{1}, \pmb{2}, \pmb{3}, \pmb{4}, \pmb{5}, \pmb{6}\}$      \\ \hline
  \end{tabular}
\label{tab:PSNR_vs_C_varying_d}
 \end{center}
 \end{table}

\begin{table*}[t]
\begin{center}
\caption{Selected subset of reference views and associated quality for scenarios with $[U_L^0,U_R^0]=[0.75,7.25]$ , $d=25$ mm, $\gamma=0.2$, $D_{max}= 200$. } \label{table:Selected_Views_NonEqSpaced}
\begin{tabular}{|c||c|c|c|c|||c|c|c|c|c|c|c|c|} \hline 
\multicolumn{5}{|c|}{$\mathcal{V}=\{0, 1, 3,  5, 7,  8\} $, case $a)$  } & \multicolumn{5}{|c|}{$\mathcal{V}=\{0, 2, 3, 4, 7, 8\}$, case $b)$ } \\ \hline  
$C $ & $\mathcal{T}_{s}$ &  PSNR &    $\mathcal{T}_{ns}$ &  PSNR   &$C$  & $\mathcal{T}_{s}$ &  PSNR    & $\mathcal{T}_{ns}$ &  PSNR        \\ \hline  \hline
$2$  & $\{0.75, 7.25\}$ &  29.39 &   $\{\pmb{0}, \pmb{8}\}$  &    28.04 &   $2$ &  $\{0.75, 7.25\}$ & 29.08  &     $\{\pmb{0}, \pmb{8}\}$     &  28.04   \\ \hline 
$3$  & $\{0.75, \pmb{3}, 7.25\}$ &  32.35  &  $\{\pmb{0}, \pmb{3}, \pmb{8}\}$  &  31.13 &  $3$ &  $\{0.75, \pmb{4}, 7.25\}$ &  32.33  &    $\{ \pmb{0},   \pmb{4},  \pmb{8} \}$ &   31.49  \\ \hline
$4$  & $\{0.75, \pmb{3}, \pmb{5}, 7.25\}$ &  35.24 &  $\{\pmb{0}, \pmb{3}, \pmb{5}, \pmb{8}\}$   &   33.87 &$4$ &  $\{\pmb{0},  \pmb{2},  \pmb{4}, 7.25\}$ &  34.18 &  $\{ \pmb{0}, \pmb{2}, \pmb{4},  \pmb{8}\}$ &  33.21   \\ \hline
$5$  & $\{\pmb{0}, \pmb{1},  \pmb{3}, \pmb{5}, 7.25\}$ &  35.85  &   $\{\pmb{0}, \pmb{1},  \pmb{3}, \pmb{5}, \pmb{8}\}$  &  35.017 & $5$ &  $\{\pmb{0},  \pmb{2},  \pmb{4}, \pmb{7}, \pmb{8}\}$ &    34.92 & $\{ \pmb{0}, \pmb{2}, \pmb{4}, \pmb{7}, \pmb{8} \}$ &    34.92     \\ \hline
$6$  & $\{\pmb{0}, \pmb{1},  \pmb{3}, \pmb{5}, \pmb{7}, \pmb{8} \}$ &   36.56 &  $\{\pmb{0}, \pmb{8}\}$   &  36.56
& $6$ &  $\{\pmb{0},  \pmb{2},  \pmb{3}, \pmb{4}, \pmb{7}, \pmb{8}\}$ &    35.60 & $\{ \pmb{0}, \pmb{2}, \pmb{4}, \pmb{7}, \pmb{8} \}$ &    35.60       \\ \hline\hline
  \end{tabular}
\end{center}
\end{table*}

We consider the cases of equally  spaced cameras $\left(\mathcal{V}=\{0,1, 2, \ldots, 5, 6\}\right)$  and unequally spaced cameras $\left(\mathcal{V}=\{0, 1, 3,  5, 7,  8\}\right.$  and  $\left.\mathcal{V}=\{0, 2, 3, 4, 7, 8\}\right)$  capturing the scene of interest.   
 In Fig. \ref{fig:PSNR_vs_C_varying_d}, we show the mean PSNR as a function of the available channel capacity $C$ when the   navigation window  requested by the user is $[0.75,5.25]$ and  cameras  are equally spaced.   The distortion of the synthesized viewpoints is evaluated with   \eqref{eq:distortion_model}, with $\gamma=0.2$, $D_I=200$, and $d= 25$.      The case of synthesis in the network is compared with the one in which only camera views can be sent to clients. In Table \ref{tab:PSNR_vs_C_varying_d}, we show the   optimal subsets $\mathcal{T}_s$ and $\mathcal{T}_{ns}$ associated to each simulation point in Fig. \ref{fig:PSNR_vs_C_varying_d}, where    camera views indexes are highlighted in bold. We observe that   the case with synthesis in the network performs best in terms of quality over the  navigation window.   When $C=2$, $\mathcal{T}_{s}:\{0.75,5.25\}$ for the network synthesis  case, and $\mathcal{T}_{ns}:\{0, 6\}$, otherwise. 
 However, the larger the channel capacity the less the need for sending virtual viewpoints. When $C=6$, for example, both camera views $0$ and $1$ can be sent, thus there is no gain in transmitting  only view $0.75$.  Finally,  when $C=7$ and all camera views can be sent to clients, $\mathcal{T}_{s} = \mathcal{T}_{ns} =\mathcal{V}$, with   $\mathcal{V}$ being the set of camera views. As expected, sending synthesized viewpoints as reference views leads to a quality gain only in constrained scenarios in which the channel capacity does not allow to send all views required for reconstructing the  navigation window  of interest.

\begin{figure}[t]
\begin{center}
\includegraphics[width=0.55\linewidth ,draft=false]{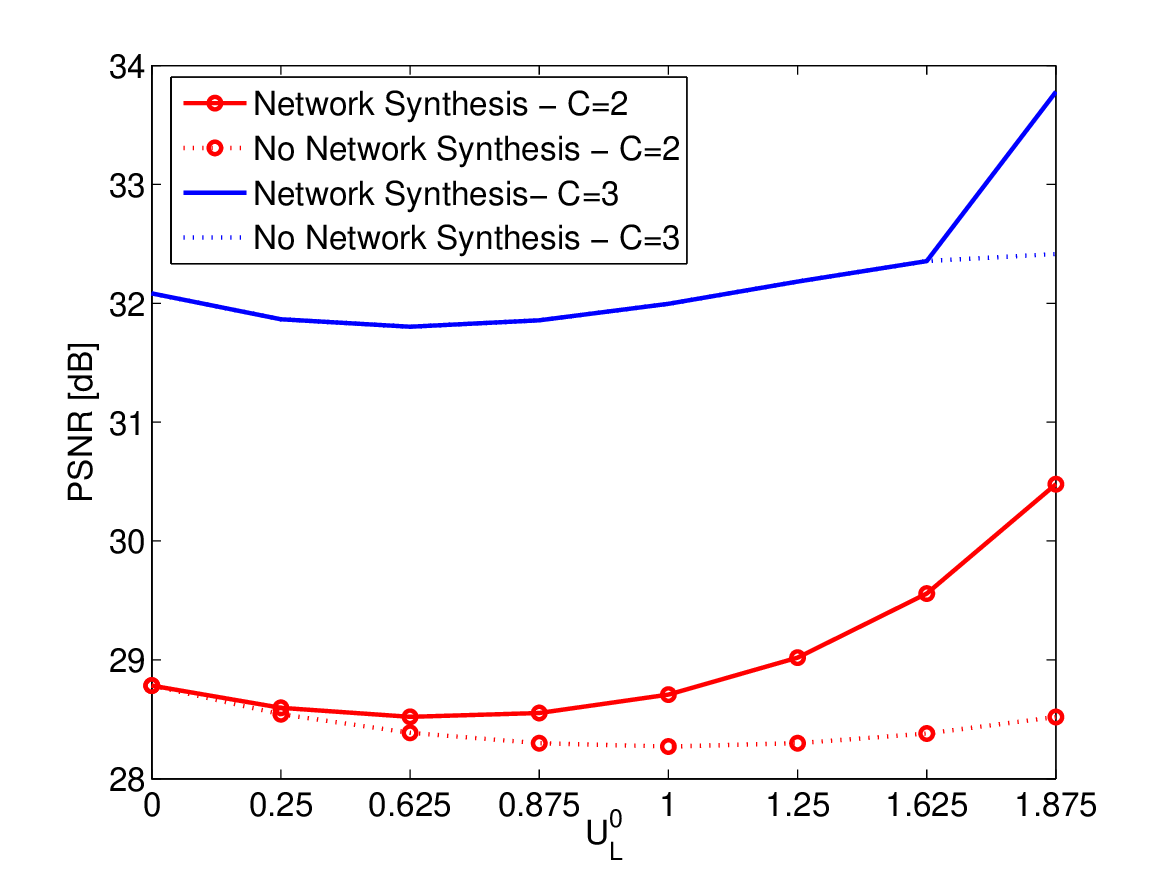} 
\caption{PSNR (in dB) vs. $U_L^0$ for a  camera set $\mathcal{V}=\{0, 2, 3, 4\}$,  navigation window    $[U_L^0, 4]$, with  $d=50$, $\gamma=0.2$, and $D_I=200$.}\label{fig:PSNR_vs_DeltaU}
\end{center}
\end{figure}

We now study the gain in allowing network synthesis when camera views are not equally spaced.  In Table \ref{table:Selected_Views_NonEqSpaced}, we provide  the optimal subsets of reference views for both sets of unequally spaced cameras $\left(\mathcal{V}=\{0, 1, 3,  5, 7,  8\}\right.$  and  $\left.\mathcal{V}=\{0, 2, 3, 4, 7, 8\}\right)$.  Similarly to the case of equally spaced cameras, we observe that virtual viewpoints are selected as reference views (i.e., they are in the best subset $\mathcal{T}_s$)  when the bandwidth $C$ is limited.   For the   camera set $a)$   the virtual view $0.75$ is selected as reference view also  for $C=4$, while the   camera set  $b)$  prefers to select the camera views $0$, $2$  at $C=4$. This is justified by the fact that in the latter scenario, the viewpoint $0.75$ is synthesized from $(V_L,V_R)=(0,2)$ thus at a larger distortion than the viewpoint $0.75$ in   scenario  $a)$, where the viewpoint is synthesized from $(V_L,V_R)=(0,1)$. This   distortion penalty makes the synthesis worthy when the channel bandwidth is highly constrained ($C=2, 3$), but not in the other cases.

 In Fig. \ref{fig:PSNR_vs_DeltaU}, the average quality of the client navigation is provided as a function of the left extreme view $U_L^0$  of the navigation window $[U_L^0, 4]$  with the  camera set $\mathcal{V}=\{0, 2, 3, 4\}$ with  $d=50$, $\gamma=0.2$, and $D_I=200$ in \eqref{eq:distortion_model}. It is worth noting that $U_L^0$ ranges from $0$ to $1.875$ and only view  $0$ is   a camera view in this range.   When $U_L^0=0$ and $C=2$, the reference views $0$ and $4$ perfectly cover  the entire navigation window requested by the user, so there is no need for sending any  virtual viewpoint as reference view. This is no more true   for $U_L^0>0$. When the channel capacity is $C=2$, the gain in allowing synthesis at the cloudlets increases with   $U_L^0$. This is justified by the fact that in a very challenging scenario (i.e., limited channel capacity), the larger   $U_L^0$ the less efficient is it is to send the reference view $0$ to reconstruct images in $[U_L^0,4]$. At the same time, sending $2$ and $4$ as reference views would not allow to reconstruct   the viewpoints lower than $2$. This gain in allowing network synthesis is reflected in the PSNR curves of Fig.  \ref{fig:PSNR_vs_DeltaU}, where we can observe    an increasing gap between the case of synthesis allowed and not allowed  for $C=2$. This gap is however reduced for the scenario of $C=3$. This is expected since the  navigation window   is a limited one, at most ranging from $0$ to $4$ and   $3$ reference views cover the navigation window pretty well.

\begin{figure}[t]
\begin{center}
\includegraphics[width=0.55\linewidth ,draft=false]{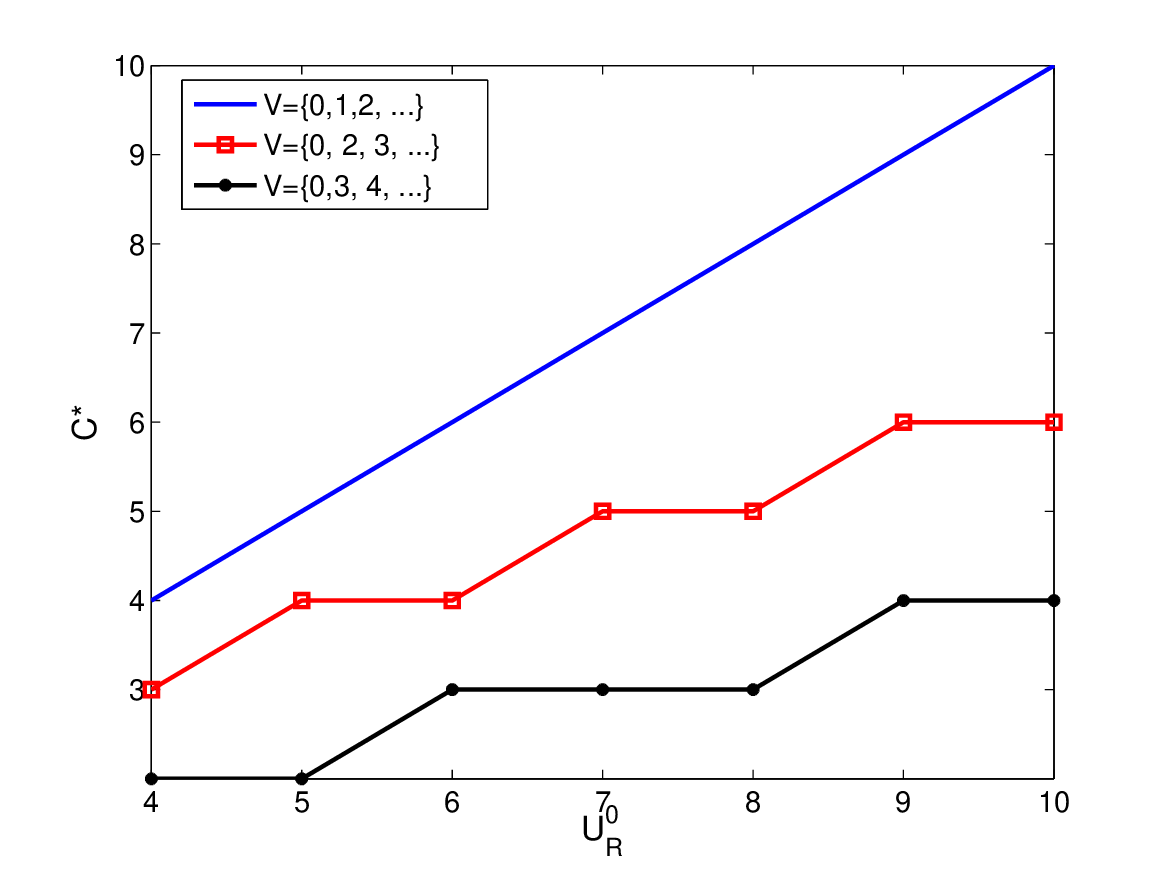} 
\caption{Thresholding $C$ vs. $U_R^0$  for a navigation window  $[0.5, U_R^0]$, with  $d=50$mm, $\gamma=0.2$, and $D_I=200$.}\label{fig:BestC_vs_UR}
\end{center}
\end{figure}
 To better show this tradeoff between distortion of the virtual reference view and  the bandwidth gain,  we introduce the  thresholding channel value, denoted by $C^{\star}$. The latter is defined as the value of channel bandwidth beyond which no gain is experienced in allowing synthesis in the network compared to a case of no synthesis. In Fig. \ref{fig:BestC_vs_UR}, we provide the behavior of the thresholding channel value as a function the navigation window, for different cameras set. In particular, we consider $U_L^0=0.5$ and we let $U_R^0$ varies from $5$ to $10$. Also, we simulate three different scenarios that differ for the available camera set. In particular, we have $\mathcal{V}= \{0,1, 2, 3, \ldots\}$, $\mathcal{V}= \{0, 2, 3, 4, \ldots\}$, and $\mathcal{V}= \{0, 3, 4, 5,  \ldots\}$. The main difference is then in the reference views that can be used to synthesize the virtual viewpoint $0.5$. In the first case, $0.5$ is reconstructed from camera views $(0,1)$ while in the last case from $(0,3)$ increasing then the distortion of the synthesis. Because of this increased distortion of $0.5$, the virtual viewpoint is not always sent as reference view. In particular, we can observe that the larger the distortion of the virtual viewpoint, the lower the thresholding channel value. This means that even in challenging scenarios, as for example in the case of $U_R^0=7$ and $C=3$, if $\mathcal{V}= \{0, 3, 4, 5,  \ldots\}$   then there is no gain in synthesize in the network, while we still have a gain if $\mathcal{V}= \{0,1, 2, 3, \ldots\}$.
 \begin{figure}[t]
\begin{center}
\includegraphics[width=0.55\linewidth ,draft=false]{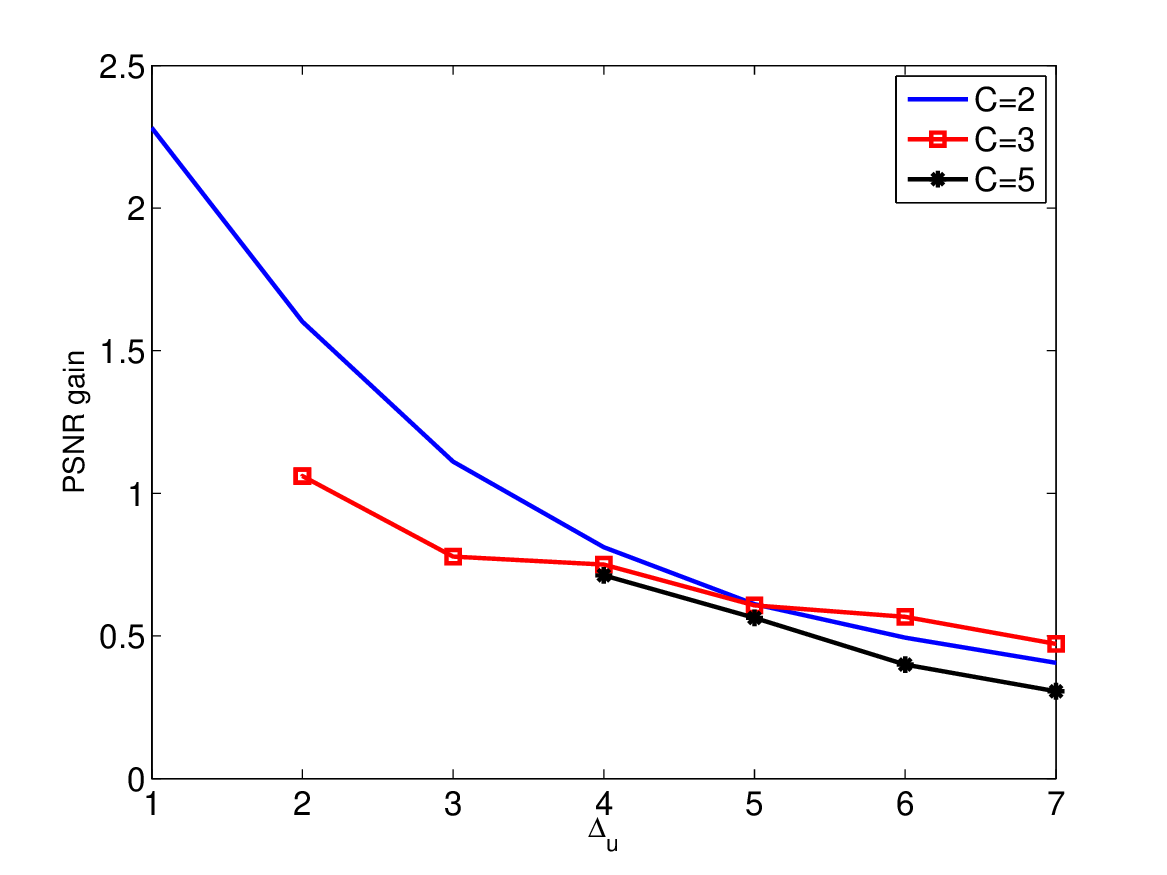} 
\caption{ PSNR gain (in dB) vs. the   navigation window   size $\Delta_u$, for different channel capacity constraints $C$ when  $d=50$mm, $\gamma=0.2$, and $D_I=200$.}\label{fig:DeltaPSNR_vs_RangeInterest}
\end{center}
\end{figure}
In Fig. \ref{fig:DeltaPSNR_vs_RangeInterest},  we provide the mean PSNR as a function of the size of the navigation window, namely $\Delta_u$. More in details, for each value of $\Delta_u$, we define the  navigation window   as $[U_L^0, U_L^0+ \Delta_u]$. The starting viewpoint $U_L^0$ is randomly selected. For each realization of the navigation window, the best subset is evaluated (both when synthesis is allowed and when it is not) and the quality of the reconstructed viewpoint   in the navigation window  is evaluated. For each  $\Delta_u$, we average the quality simulating all possible $U_L^0$ starting viewpoint within a total range of $[0,12]$. In the results we provide the PSNR gain, defined as the difference between the mean PSNR (in dB) when the synthesis is allowed and the  mean PSNR (in dB) when only  camera views are considered as reference views. Thus, the figure shows the gain in synthesizing for different sizes of the navigation window. As general trend, we observe that the quality gain decreases with  $\Delta_u$. This is due to the fact that the gain mainly comes from the lateral reference views, that are usually virtual viewpoints if synthesis is allowed. This leads to a gain that is however reduced for large sizes of the navigation window. 
Finally, we also observe that the gain does not necessarily depends on the channel constraint $C$.

\begin{figure}[t]
\begin{center}
\includegraphics[width=0.55\linewidth ,draft=false]{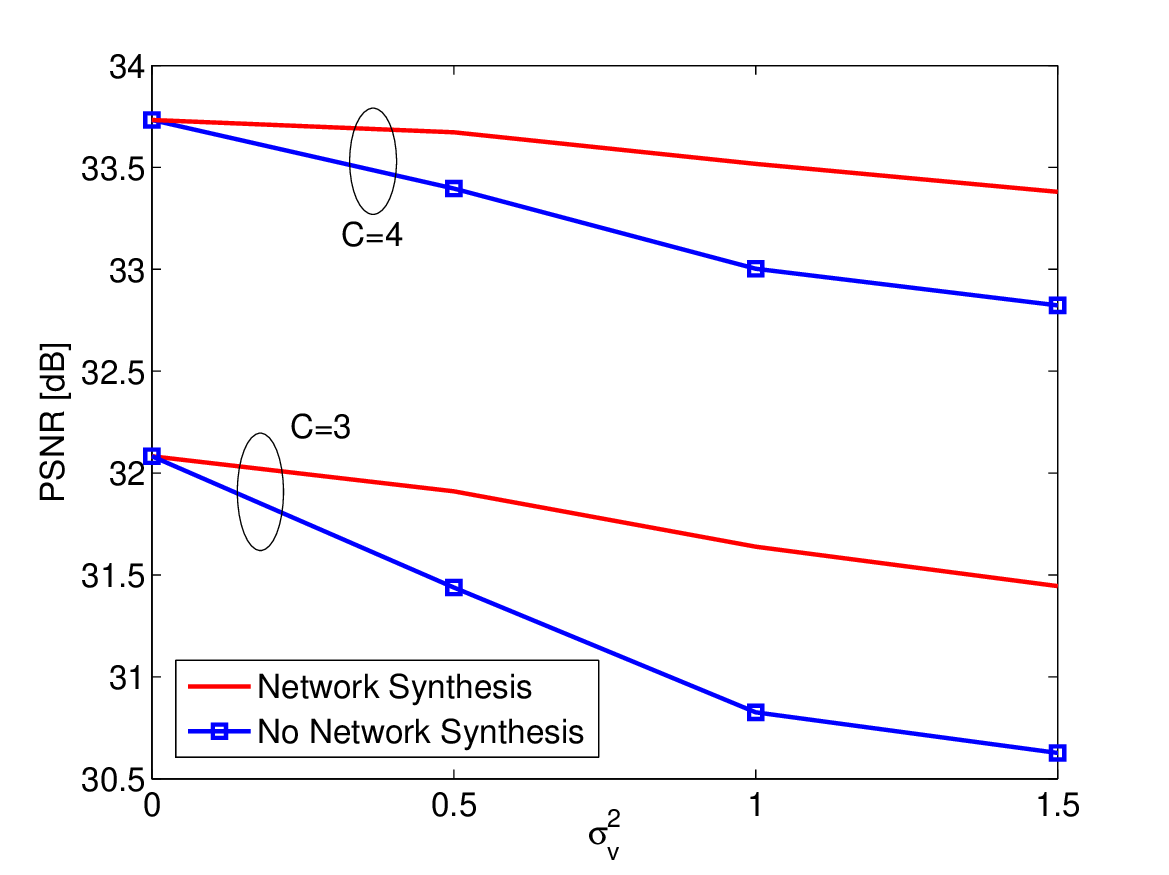} 
\caption{ PSNR  (in dB) vs.   $\sigma_v^2$, for different channel capacity constraints $C$ when  $d=50$, $\gamma=0.2$, and $D_I=200$.}\label{fig:Q_vs_Randomness}
\end{center}
\end{figure}

We now   consider a scenario in which the camera views position is not a priori given. In Fig. \ref{fig:Q_vs_Randomness}, we provide the mean PSNR as a function of the variance $\sigma_v^2$, which defines the randomness  of the camera views positions when acquiring the scene. More in details, we consider a navigation window  $[U_L^0,U_R^0]=[2,6]$. We then define a deterministic camera views set $\mathcal{V}_D=\{0,1,2, \ldots, 6, 7\}$, which is the best camera view set since it is aligned with the requested viewpoint navigation window. For each value of $\sigma_v^2$, we generate a random cameras set  $\mathcal{V}$ as $\mathcal{V} =\mathcal{V}_D + [n_0, n_1, \ldots, n_7]$, where each $n_i$ is a gaussian random variable with zero mean and variance $\sigma_v^2$ with $n_i$ and $n_j$ mutually independent for $i\neq j$. Thus, the larger $\sigma_v^2$, the larger the probability for the camera view set  to be not aligned with the  navigation window. For each realization of $\mathcal{V}$, we run our optimization for both the cases of allowed and not allowed synthesis and we evaluate the experienced quality. For each $\sigma_v^2$ value we simulate $400$ runs and we provide in  Fig. \ref{fig:Q_vs_Randomness} the averaged quality. 
What it is interesting to observe is that even if camera views are not perfectly aligned with the navigation window  of interest (i.e., even for large variance values) the quality degradation with respect to the case of $\sigma_v^2=0$ is limited, about $0.5$ dB for $C=3$, when network synthesis is allowed. On the contrary, when synthesis is not allowed in the cloudlet, the quality substantially decreases with $\sigma_v^2$, experiencing a PSNR loss of almost $1.5$dB.  This means that network synthesis can compensate  for cameras not ideally positioned in the 3D scene,  as in the case of  user generated content systems.

\begin{figure}[t]
\begin{center}
\includegraphics[width=0.55\linewidth ,draft=false]{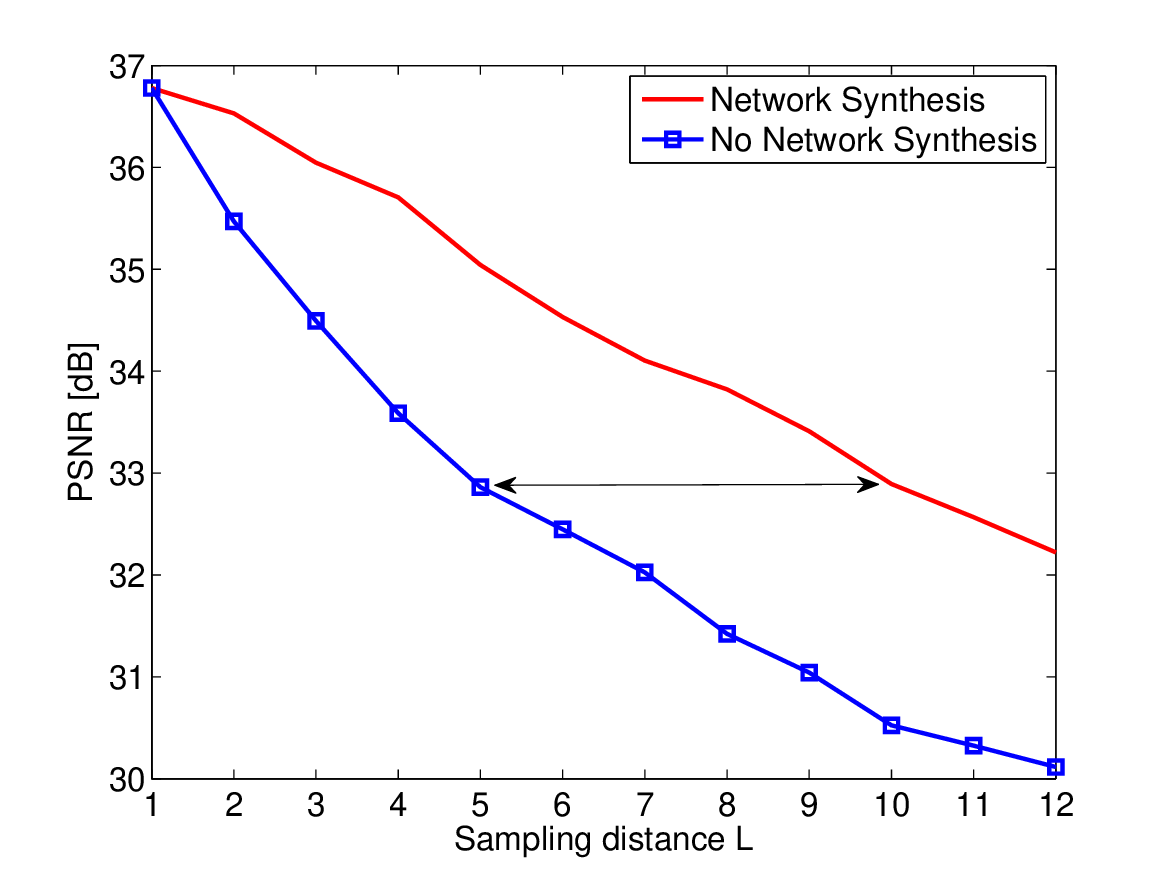} 
\caption{ PSNR   (in dB) vs.  the sampling distance $L$,  for $C=2$,   $d=50$, $\gamma=0.2$, and $D_I=200$.}\label{fig:Q_vs_Cameras}
\end{center}
\end{figure}

Finally, we study performance of the cloudlet-based view synthesis for a varying number of acquiring cameras. In particular, given the set of equally spaced viewpoints $\mathcal{U}$,  we assume that one   every $L$ viewpoints in $\mathcal{U}$ is a camera view, i.e.,  there are $L-1$ virtual viewpoints between consecutive camerasview. Being the viewpoints in $\mathcal{U}$ equally spaced, say at distance $d$,  $L d$ is the distance between consecutive cameras. 
In the following,  we provide the quality behavior for $L$ ranging from $1$ to $12$.  For each value of the sampling distance $L$, we simulate a navigation window  spanning a range of $20 d$. The navigation window is    selected uniformly at random and the optimization algorithm evaluates the best subset of reference views. The experienced quality is averaged over $400$ runs and evaluated for different values of  $L$.  
In Fig. \ref{fig:Q_vs_Cameras}, we show the mean quality for the navigation as a function of the sampling distance $L$, for the scenario with  for $C=2$,   $d=50$, $\gamma=0.2$, and $D_I=200$ in \eqref{eq:distortion_model}. It is worth noting that for a user to navigate at given quality, a much higher value of sampling distance $L$ can be used when  network synthesis is allowed, with respect to the value of $L$ required with no network synthesis. For example, a mean quality in the navigation of $33$ dB is achieved with $L=5$  when network synthesis is not allowed as opposed to   $L=10$ when allowing network synthesis. This means that   when synthesis is allowed, half of the number of camera views can be used respect to the case in which no synthesis is allowed. Thus,  view synthesis  in the network   allows to maintain a good navigation quality when reducing the number of cameras.